\documentclass{ifacconf}
\usepackage{graphicx}      
\usepackage{natbib}        

\usepackage[utf8]{inputenc}
\usepackage{amsfonts,amsmath}
\usepackage{mdframed}
\usepackage{graphicx}
\usepackage{hyperref}

\usepackage{multirow}
\usepackage{array}
\usepackage{subcaption}

\usepackage{url}

\usepackage{tikz}
\usetikzlibrary{arrows.meta,,positioning}

\usepackage{caption}

\def\smskip{\vskip 5 pt}
\def\medskip{\vskip 10 pt}
\def\bigskip{\vskip 15 pt}
\def\pn{\par\noindent}
\def\br{\break}

\def\bl{\bigl} 
\def\br{\bigr}

\def\tl{\tilde}

\def\old#1{ }

\def\tends{\rightarrow}

\def\frac#1#2{{#1\over #2}}

\def\ol#1{\overline{#1}}

\def\a{\alpha}

\def\m{\mu}

\begin{document}
\begin{frontmatter}

\title{Model Predictive Control and Reinforcement Learning:\break A Unified Framework Based on Dynamic Programming} 

\author[First]{Dimitri P.\ Bertsekas} 

\address[First]{Arizona State University, Tempe, AZ  USA}

\begin{abstract}  
In this paper we describe a new conceptual framework that connects approximate Dynamic Programming (DP), Model Predictive Control (MPC), and Reinforcement Learning (RL). This framework centers around two algorithms, which are designed largely independently of each other and operate in synergy through the powerful mechanism of Newton's method. We call them the {\it off-line training} and the {\it on-line play} algorithms. The names are borrowed from some of the major successes 
of RL involving games; primary examples are the recent (2017) AlphaZero program (which plays chess, [SHS17], [SSS17]), and the similarly structured and earlier (1990s) TD-Gammon program (which plays backgammon, [Tes94], [Tes95], [TeG96]). In these game contexts, the off-line training algorithm is the method used to teach the program how to evaluate positions and to generate good moves at any given position, while the on-line play algorithm is the method used to play in real time against human or computer opponents.

Significantly, the synergy between off-line training and on-line play also underlies MPC (as well as other major classes of sequential decision problems), and indeed the MPC design architecture is very similar to the one of AlphaZero and TD-Gammon. This  conceptual insight provides a vehicle for bridging the cultural gap between RL and MPC, and sheds new light on some fundamental issues in MPC. These include the enhancement of stability properties through rollout, the treatment of uncertainty through the use of certainty equivalence, the resilience of MPC in adaptive control settings that involve changing system parameters, and the insights provided by the superlinear performance bounds implied by Newton's method.
\vspace{0.5pc} 

{\bf To be published in Proc.\ of IFAC NMPC, Kyoto, August 2024}
\end{abstract}


\end{frontmatter}

\section{Introduction}
We will describe a conceptual framework for approximate DP, RL, and their connections to MPC, which was first presented in the author's recent books [Ber20] and [Ber22a]. The present paper borrows heavily from these books, the course textbook [Ber23], the overview papers [Ber21a], [Ber22c], as well as recent research by the author and his collaborators.\footnote{Special thanks are due to Yuchao Li for extensive helpful interactions relating to many of the topics discussed in this paper. Early discussions on MPC with Moritz Diehl were greatly appreciated. The suggestions of Manfred Morari and James Rawlings, as well as those of the reviewers, were also very much appreciated.}

Our framework is very broadly applicable thanks to the generality of the DP methodology on which it rests. This generality allows arbitrary state and control spaces, thus facilitating a  free movement between continuous-space infinite-horizon formulations (such as those arising in control system design and MPC), discrete-space finite-horizon problem formulations (such as those arising in games and integer programming), and mixtures thereof that involve both continuous and discrete decision variables.

To present our framework, we will first focus on a class of deterministic discrete-time optimal control problems, which underlie typical MPC formulations. In subsequent sections, we will indicate how the principal conceptual components of our framework apply to problems that involve stochastic as well as set membership uncertainty, and how they impact the effectiveness of MPC for indirect adaptive control.

\subsection{An MPC Problem Formulation}

The theory and applications of MPC has undergone extensive development, since the early days of optimal control, thanks to research efforts from several scientific communities.\footnote{The idea underlying MPC is on-line optimization with a truncated rolling horizon and a terminal cost function approximation. This idea has arisen in several contexts, motivated by different types of applications. It has been part of the folklore of the optimal control  and operations research literature, dating to the 1960s and 1970s. Simultaneously, it was used in important chemical process control applications, where the name ``model predictive control" (or ``model-based predictive control") and the related name ``dynamic matrix control" were introduced. The term ``predictive" arises often in this path breaking literature, and generally refers to taking into account the system's future, while applying control in the present. Related ideas appeared independently in the computer science literature, in contexts of search ($A^*$ and related), planning, and game playing.}The early papers by Clarke, Mohtadi, and Tuffs [CMT87a], [CMT87b], Keerthi and Gilbert [KeG88], and Mayne and Michalska [MaM88], attracted significant attention. Surveys, which give many of the early references, were given by Morari and Lee [MoL99], Mayne et al.\ [MRR00], Findeisen et al.\ [FIA03], and Mayne [May14]. Textbooks such as Maciejowski [Mac02], Goodwin, Seron, and De Dona [GSD06], Camacho and Bordons [CaB07], Kouvaritakis and Cannon [KoC16], Borrelli, Bemporad, and Morari [BBM17], Rawlings, Mayne, and Diehl [RMD17], and Rakovic and Levine [RaL18], collectively provide a comprehensive view of the MPC methodology. 

More recent works have aimed to integrate ``learning" into MPC, similar to the practices of the RL and AI communities. This line of research is very active at present; for some representative papers, see [CLD19], [GrZ19], [Rec19], [CFM20], [HWM20], [MGQ20], [BeP21], [KRW21], [BGH22], [CWA22],  [GrZ22], [MDT22], [MJR22], [SKG22], and [DuM23]. 

To provide an overview of the main ideas of our framework, let us consider a deterministic stationary discrete-time system of the form
$$x_{k+1}=f(x_k,u_k),\qquad k=0,1,\ldots,$$
where $x_k$ and $u_k$ are the state and control at time $k$, taking values in some spaces $X$ and $U$.
We consider stationary feedback policies $\m$, whereby at a state $x$ we apply control $u=\m(x)$, subject to the constraint that $\m(x)$ must belong to a given set  $U(x)$ for each $x$. 

The cost function of $\m$, starting from an initial state $x_0$ is 
 $$J_\m(x_0) =
\lim_{N\tends\infty}\sum_{k=0}^{N-1}
\a^kg\bl(x_k,\mu(x_k)\br),$$
where $\a\in(0,1]$ is a discount factor, and
$$g(x,u)\ge 0,\qquad \hbox{for all }x\in X, u\in U(x).$$
We also assume that there is a cost-free and absorbing termination state $t$  [i.e., $g(t,u)=0$ and $f(t,u)=t$ for all $u\in U(t)$]; e.g., the origin in typical optimal regulation settings in control.
The optimal cost function is defined by
$$J^*(x)=\min_{\m\in{\cal M}}J_\m(x),\qquad \forall\ x\in X,$$
where ${\cal M}$ is the set of all admissible policies, and our objective is to find an optimal policy $\m^*$, i.e., one that satisfies $J_{\m^*}(x)=J^*(x)$ for all $x\in X$. 

This is a typical MPC problem formulation, and it includes the classical linear-quadratic problems where $X=\Re^n$, $U=\Re^m$, $f$ is linear, $g$ is positive definite quadratic, and the termination state $t$ is the origin of $\Re^n$.  
Note that our formulation makes no assumptions on the nature of the state and control spaces $X$ and $U$; they can be arbitrary.  However, the problem and its computational solution have been analyzed at the level of generality used here in the author's paper [Ber17b], which can serve as a foundation for mathematical results and analysis that we will use somewhat casually in this paper.

Stability of policies is of paramount importance in MPC. In particular, the issue of stability was addressed theoretically by
Keerthi and Gilbert [KeG88], and stability issues have been discussed in detail in the overview paper by Mayne et al.\ [MRR00]. A stability analysis with discrete constraint sets was given by Rawlings and Risbeck [RaR17]. The  paper by Krener [Kre19] considers methods to estimate the optimal cost function for use as terminal cost function, aiming to achieve stabilization with MPC lookahead that is as small as possible. 

In the context of the present paper, however, because $X$ and $U$ can be arbitrary sets, it is necessary to use a more general line of analysis and a nontraditional definition of stability. In particular, we say that a policy $\m$ is {\it stable} if 
$$J_{\m}(x)<\infty,\qquad  \forall\ x\in X.$$
For problems where $\a=1$, this definition of stability is qualitatively similar to traditional definitions of stability in control theory/MPC contexts, including linear-quadratic problems (to be used later for visualization purposes). Our subsequent discussion of stability implicitly assumes such a context, and may not be meaningful in other contexts, such as games, discrete optimization, cases where $\a<1$, etc.
Note that $J^*(x)$ is finite for all $x$ if there exists at least one stable policy, which we will  assume in this paper. \old{Alternatively, $J^*(x)$ is finite for all $x$ if the cost per stage $g(x,u)$ is bounded over $(x,u)$ and $\a<1$, the ``easy" discounted case of infinite horizon DP, where all policies are stable according to our definition of stability.}

\subsection{Approximation in Value Space - MPC and RL}

It is known that $J^*$ satisfies the Bellman equation
$$J^*(x)=\min_{u\in U(x)}\Big\{g(x,u)+\a J^*\big(f(x,u)\big)\Big\},\qquad \forall\ x\in X,$$ 
and that if $\m^*(x)$ attains the minimum above for all $x$, then $\m^*$ is an optimal policy. Moreover for a policy $\m$, we have
$$J_{\m}(x)=g\big(x,\m(x)\big)+\a J_{\m}\Big(f\big(x,\m(x)\big)\Big),\qquad \forall\ x\in X.$$ 
These are results that are generally accepted in the optimal control literature. Their rigorous mathematical proofs at the level of generality considered here are given in the paper [Ber17b], which relies on the general theory of abstract DP problems with nonnegative cost, developed in the paper [Ber77] and extensively discussed in the books [BeS78], [Ber22b]; see also Ch.\ 3 of the thesis [Li23] for a related discussion.

\begin{figure}
\begin{center}
\includegraphics[width=7.0cm]{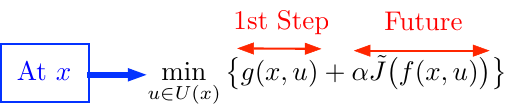}    
\caption{Illustration of approximation in value space with one-step lookahead.} 
\label{figapproxvaluespace}
\end{center}
\end{figure}

 A major RL approach, which we call {\it approximation in value space\/}, is to replace $J^*$ with an approximating real-valued function  $\tl J$, and obtain a suboptimal policy $\tl \m$ with the minimization 
$$\tl \m(x)\in\arg\min_{u\in U(x)}\big\{g(x,u)+\a \tl J\big(f(x,u)\big)\big\},\qquad \forall\ x\in X;$$
see Fig.\ \ref{figapproxvaluespace}. We assume that the minimum above is attained for all $x\in X$, and refer to $\tl \m$ as the {\it one-step lookahead policy\/}.

\begin{figure}
\begin{center}
\includegraphics[width=8.4cm]{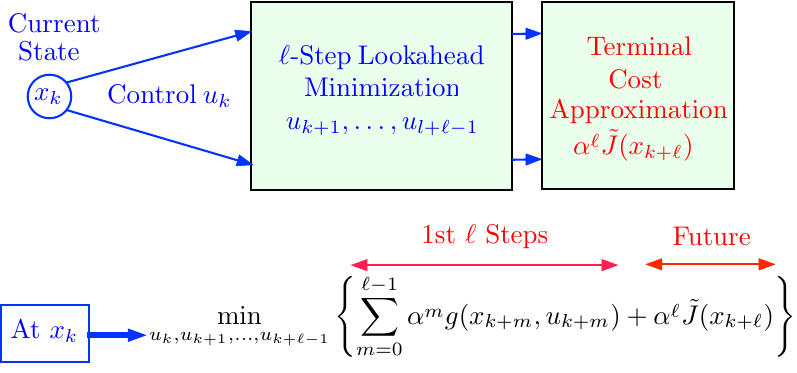}    
\caption{Illustration of approximation in value space with $\ell$-step lookahead. The $\ell$-step minimization at $x_k$ yields a sequence $\tl u_k,\tl u_{k+1},\ldots,\tl u_{k+\ell-1}$. The control $\tl u_k$ is applied at $x_k$, and defines the $\ell$-step lookahead policy $\tl \m$ via $\tl \m(x_k)=\tl u_k$. The controls $\tl u_{k+1},\ldots,\tl u_{k+\ell-1}$ are discarded. This is similar to mainstream MPC schemes.
\vspace{5 pt} }
\label{figellapproxvaluespace}
\end{center}
\end{figure}

There is also an $\ell$-step lookahead version of the preceding approach, which involves the solution of an $\ell$-step DP problem, where $\ell$ is a positive integer, with a terminal cost function approximation $\tl J$. Here at a state $x_k$ we minimize the cost of the first $\ell$ stages with the future costs approximated by $\tl J$ (see Fig.\ \ref{figellapproxvaluespace}). If this minimization yields a control sequence $\tl u_k, \tl u_{k+1},\ldots,\tl u_{k+\ell-1}$, we apply the control $\tl u_k$ at $x_k$, and discard the controls $\tl u_{k+1},\ldots,\tl u_{k+\ell-1}$. This defines a policy $\tl \m$ via $\tl \m(x_k)=\tl u_k$.

Actually, we may view $\ell$-step lookahead minimization as the special case of its one-step counterpart where the lookahead function is the optimal cost function of an $(\ell-1)$-stage DP problem that starts at $x_{k+1}$ and has a terminal cost $\a^\ell \tl J(x_{k+\ell})$ after $\ell-1$ stages.

Note that the multistep scheme depicted in Fig.\ \ref{figellapproxvaluespace} can be recognized as the most common MPC architecture design (usually $\a=1$ is chosen in MPC). When the $\ell$-step lookahead minimization problem involves continuous control variables, this minimization can often be done by nonlinear programming algorithms, such as sequential quadratic programming and related methods; for some representative papers, see [ABQ99], [WaB10], [OSB13], [BBM17], [RMD17], [LHK18], [Wri19], [FXB22]. However, when discrete/integer variables are involved, time consuming mixed integer programming computations or space and control discretization methods may be required [BeM99], [BBM17].

In MPC problems that involve state constraints, it may  also be necessary to modify the state space $X$ to ensure that the $\ell$-step lookahead minimization has a feasible solution (i.e., that the control can keep the state within $X$). This leads to the problem of {\it reachability of a target tube\/}, which was first formulated and analyzed in the author's PhD thesis [Ber71] and papers [BeR71], [Ber72], and subsequently discussed and adapted more broadly in the control and MPC literature, e.g., [KoG98], [Bla99], [Ker00], [RKM06], [GFA11], [May14], [CLL23], and [XDS23]. In the context of the off-line training/on-line play conceptual framework of the present paper, reachability issues are ordinarily  dealt with off-line, as they tend to involve substantial preliminary target tube calculations.  An alternative and simpler possibility is to replace the state constraints with penalty or barrier functions as part of the cost per stage.

Several RL methods are available for computing suitable terminal cost approximations $\tl J$ by using some form of learning from data, thus circumventing the solution of Bellman's equation. The approximation in value space approach has also received a lot of attention in the MPC literature, but in the early days of MPC there was little consideration of learning that involves training of neural networks and other approximation architectures, as practiced by the RL community.

\subsection{Rollout with a Stable Policy}

An important cost function approximation approach is rollout, where $\tl J$ is the cost function $J_{\m}$ of a stable policy $\m$, i.e., one for which $J_{\m}(x)<\infty$ for all $x\in X$. We discuss this approach in this section, together with associated stability issues.

In the MPC context it is often critical that the policy $\tl \m$ obtained by one-step and $\ell$-step lookahead is stable. It can be shown that $\tl \m$ is stable if $\tl J$ satisfies the following version of a Lyapunov condition:
$$\tl J(x)\ge \min_{u\in U(x)}\big\{g(x,u)+\a \tl J\big(f(x,u)\big)\big\},\qquad \forall\ x\in X;$$
see [Ber17b], [Ber20]. In particular, 
if $\tl J=J_\m$ for some stable policy $\m$, then $J_\m$ is real-valued and satisfies the preceding Lyapunov condition.\footnote{Note that if $\m$ is unstable, then $J_\m$ is not real-valued and does not qualify for use as $\tl J$ in the one-step lookahead scheme.} To see this, note that  from Bellman's equation we have,
$$J_\m(x)=g\big(x,\m(x)\big)+\a J_\m\Big(f\big(x,\m(x)\big)\Big),$$
so that
$$J_\m(x)\ge \min_{u\in U(x)}\big\{g(x,u)+\a J_{\m}\big(f(x,u)\big)\big\},$$
for all $x\in X$. Thus $J_\m$ satisfies the Lyapunov condition, implying that $\tl\m$ is stable when $\tl J=J_{\m}$. In this case we call $\m$ the {\it base policy\/}, and we call $\tl\m$ the {\it rollout policy that is based on $\m$\/}.

Rollout is a major RL approach, which is simple and very reliable, based on extensive computational experience. It is closely connected to the MPC design philosophy, as has been discussed in the author's early overview paper [Ber05a] and recent books. An important conceptual point is that rollout consists of a single iteration of the fundamental DP method of policy iteration, whose connection with Newton's method in the context of linear-quadratic problems [BeK65], [Kle68], and other Markov decision  problems [PoA67], [PuB78], [PuB79]  is well known.  

The main difficulty with rollout is that computing the required values of $J_\m\big(f(x,\m(x))\big)$ on-line may require time consuming simulation. This is an even greater difficulty for the $\ell$-step lookahead version of rollout, where the required number of base policy values increases exponentially with $\ell$.  In this case, approximate versions of rollout may be used, such as {\it simplified rollout\/}, {\it truncated rollout\/}, and {\it multiagent rollout\/}; see the books [Ber19], [Ber20], [Ber22a], [Ber23], and the subsequent discussion.

 \vspace{-3pt}

\subsection{Off-Line Training and On-line Play}

Implicit in approximation in value space is a conceptual separation between two algorithms:
\begin{itemize}
\item[(a)] The {\it off-line training} algorithm, which designs the cost function approximation $\tl J$, and possibly other problem components (such as for example a base policy for rollout, or a target/safety tube of states where the system must stay at all times). 
\smskip
\item[(b)] The {\it on-line play} algorithm, which implements the policy $\tl \m$ in real-time via one-step or $\ell$-step lookahead minimization, cf.\ Fig.\ \ref{figellapproxvaluespace}. 
\end{itemize}

An important point is that the off-line training and on-line play algorithms can often be designed independently of each other. In particular, approximations used in the on-line lookahead minimization need not relate to the methods used for construction of the terminal cost approximation $\tl J$. Moreover, $\tl J$ can be simple and primitive, particularly in the case of multistep lookahead, or it  may be based on sophisticated off-line training methods involving neural networks. 

Alternatively,  $\tl J$ may be computed off-line with a {\it problem approximation approach\/}, as the optimal or nearly optimal cost function of a simplified optimization problem, which is more convenient for computation (e.g., a linear-quadratic problem approximation, following linearization of nonlinear dynamics of the original problem). Problem simplifications may include exploiting decomposable structure, reducing the size of the state space, neglecting some of the constraints, and ignoring various types of uncertainties.\footnote{Two successful applications of problem approximation exploiting decomposable structures, where the author was personally involved, are described in the papers [KGB82] and [MLW24]. Another type of problem approximation, involving the use of some type of certainty equivalence, will be discussed in Section 3.} 

We note that the off-line training/on-line play separation does not explicitly appear in early MPC frameworks, but it is often used in more recent MPC proposals, noted earlier,  where $\tl J$ may involve the training of neural networks with data. On the other hand, the off-line training/on-line play division is common in RL schemes, as well as game programs such as computer chess and backgammon, which we discuss in the next section.

\subsection{AlphaZero and TD-Gammon}

The development of the AlphaZero program  by DeepMind Inc, as described in the papers [SHS17], [SSS17], is perhaps the most impressive success story in reinforcement learning (RL) to date. AlphaZero plays Chess, Go, and other games, and is an improvement in terms of performance and generality over the earlier AlphaGo program [SHM16], which plays the game of Go only. AlphaZero plays chess and Go as well or  better than all competitor computer programs, and much better than all humans.

The AlphaZero program is remarkable in several other ways. In particular, it has learned how to play without human instruction, just  data generated by playing against itself. In RL this is called {\it self-learning\/}, and can be viewed as  a form of the classical DP method of policy iteration, adapted to off-line training with self-generated data. Moreover, AlphaZero learned how to play chess very quickly; within hours, it played better than all humans and computer programs (with the help of awesome parallel computation power, it must be said). 

  \begin{figure}
\begin{center}
\includegraphics[width=8.0cm]{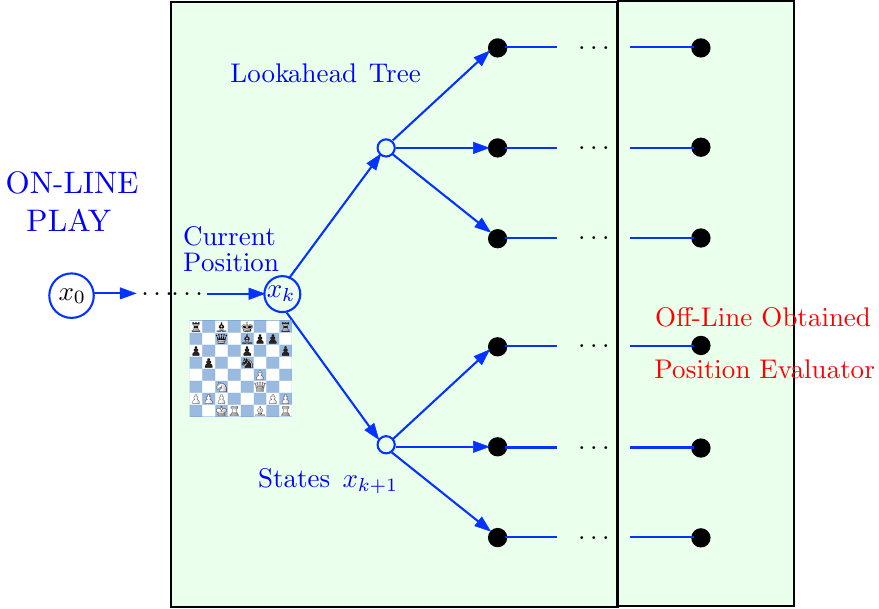}    
\caption{Illustration of the architecture of AlphaZero chess. It uses a very long lookahead minimization involving moves and countermoves of the two players followed by a terminal position evaluator, which is designed through extensive off-line training using a deep neural network. There are many implementation details that we will not discuss here; for example the lookahead is selective, because some lookahead paths are pruned, by using a form of Monte Carlo tree search. Also a primitive form of rollout is used at the end of the lookahead minimization to resolve dynamic terminal positions. Note that the off-line-trained  neural network of AlphaZero produces both a position evaluator and a playing policy. However, the neural network-trained policy is not used directly for on-line play.} 
\label{figchess_IFAC}
\end{center}
\end{figure}

The architecture of  AlphaZero is described in Fig.\ \ref{figchess_IFAC}. A comparison with Fig.\ \ref{figellapproxvaluespace} shows that the architectures of AlphaZero and MPC are very similar. They both involve optimization over a truncated rolling horizon with a terminal cost approximation.\footnote{
Note that  AlphaZero is trained to select moves assuming that it plays against an adversarial opponent. Its design philosophy would be more closely aligned to MPC, if it were to play against a known and fixed opponent, whose moves can be perfectly predicted at any given position.}
In AlphaZero, the cost function approximation takes the form of a position evaluator, which uses a deep neural network, trained off-line with an immense amount of chess data. The neural network training process also yields a player that can select a move ``instantly" at any given chess position, and can be used to assist the on-line lookahead process.

The success of the AlphaZero design framework was replicated by other chess programs such as LeelaChess and Stockfish. It is presently believed that the principal contributor to their success is long lookahead, which uses an efficient on-line play algorithm that involves various forms of tree pruning.
The off-line trained position evaluator and player have also contributed to success, although likely to a lesser extent.

 \begin{figure}
\begin{center}
\includegraphics[width=8.4cm]{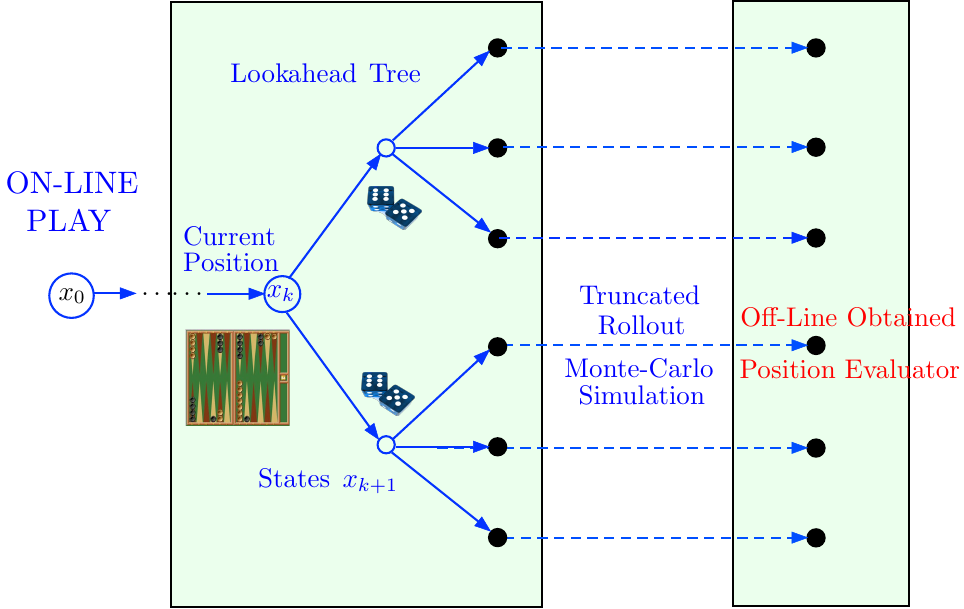}    
\caption{Illustration of the architecture of TD-Gammon with truncated rollout [TeG96]. It uses a relatively short lookahead minimization followed by rollout and terminal  position evaluation (i.e.,  game simulation with the one-step lookahead player; the game is truncated after a number of moves, with a position evaluation at the end). Note that backgammon involves stochastic uncertainty, and its state is the pair of current board position and dice roll.} 
\label{figBackgammon_IFAC}
\end{center}
\end{figure}

The principles of the AlphaZero design have much in common with the earlier TD-Gammon programs of Tesauro [Tes94], [Tes95], [TeG96] that play backgammon (a game of substantial computational and strategical complexity, which involves a large state space, as well as stochastic uncertainty due to the rolling of dice); see Fig.\ \ref{figBackgammon_IFAC}. TD-Gammon also uses an off-line neural network-trained terminal position evaluator, and importantly, in its 1996 version, it also extends its on-line lookahead by rollout. Tesauro's programs stimulated much interest in RL in the middle 1990s, and the last of these programs [TeG96] exhibits different and better play than humans. The rollout algorithm, based on Monte-Carlo simulation, has been a principal contributor to this achievement.\footnote{The name ``rollout" was coined by Tesauro [TeG96] in the context of backgammon. It refers to simulating/``rolling out" and averaging the scores of many backgammon games, starting from the current position and using the one-step lookahead player that is based on the  position evaluator.}

A striking empirical observation is that while the neural network used in AlphaZero was trained extensively, the player that it obtained off-line is not used directly during on-line play (it is too inaccurate due to approximation errors that are inherent in off-line neural network training). Instead a separate on-line player is used to select moves, based on multistep lookahead minimization, a limited form of rollout, and a terminal position evaluator that was trained using experience with the off-line player (cf.\ Fig.\ \ref{figchess_IFAC}). The on-line player performs a form of policy improvement, which is not degraded by neural network approximations. As a result, it greatly improves the performance of the off-line player. 

Similarly, TD-Gammon performs on-line a policy improvement step using one-step or two-step lookahead minimization, which is not degraded by neural network approximations. Note that the lookahead minimization in computer backgammon is short, because  its lookahead tree of moves and countermoves expands very quickly to take into account the stochastic dice rolls. However, rollout with a base policy, aided by a trained neural network that provides position evaluations, effectively expands the length of the lookahead.   

Thus in summary:

\begin{itemize}
\item[(a)] The on-line player of AlphaZero plays much better than its extensively trained off-line player. This is due to the beneficial effect of  policy improvement with long lookahead minimization, which corrects for the inevitable imperfections of the neural network-trained off-line player, and position evaluator/terminal cost approximation.
\smskip
\item[(b)] The TD-Gammon player that uses long rollout with a policy plays much better than TD-Gammon without rollout. This is due to the beneficial effect of the rollout, which serves as a substitute for long lookahead minimization.  	
\end{itemize}

An important lesson from AlphaZero and TD-Gammon is that the performance of an off-line trained policy can be greatly improved by on-line approximation in value space, with long lookahead (involving minimization or rollout with the off-line policy, or both), and terminal cost approximation that is obtained off-line.
This performance enhancement is often dramatic and is due to a simple fact, which is couched on algorithmic mathematics and is a focal point of the present paper: 

\begin{itemize}
\item[(a)] {\it Approximation in value space with one-step lookahead minimization amounts to a step of Newton's method for solving Bellman's equation\/}.
\smskip
\item[(b)] {\it The starting point for the Newton step is based on the results of off-line training, and can be enhanced by longer lookahead minimization and on-line rollout\/}.  	
\end{itemize}

Indeed the major determinant of the quality of the on-line policy is the Newton step that is performed on-line, while off-line training plays a secondary role by comparison.

\subsection{An Overview of our Framework}

In the next section, we will aim to illustrate the principal ideas of our framework. These are the following:
\begin{itemize}
\item[(a)] One-step lookahead is equivalent to a step of Newton's method for solving the Bellman equation.
\item[(b)] $\ell$-step lookahead is equivalent to a step of a Newton/SOR method, whereby the Newton step is preceded by $\ell-1$ SOR steps (a form of DP/value iterations; SOR stands for {\it successive over-relaxation} in numerical analysis terminology). 
\item[(c)] There is a superlinear relation between the approximation error $\|\tl J-J^*\|$ and the performance error  $\|J_{\tl\m}-J^*\|$, owing to the preceding Newton step interpretation. As a result, within the region of convergence of Newton's method, the performance error $\|J_{\tl\m}-J^*\|$ is small and often negligible. In particular,  the MPC policy ${\tl\m}$ is very close to optimal if $\tl J$ lies within the region of superlinear convergence of Newton's method.
\item[(d)] The region of convergence of Newton's method expands as the length $\ell$ of the lookahead minimization increases. Thus the performance of the MPC policy $\tl \m$ improves as $\ell$ increases, and is essentially optimal if $\ell$ is sufficiently large {\it regardless of the quality of the terminal cost approximation $\tl J$\/}.  Indeed, for finite state and control spaces, discount factor $\a<1$, and a long enough lookahead, it can be shown that {\it $\tl \m$ is an optimal policy, regardless of the size of  the approximation error $\|\tl J-J^*\|$\/}; see Appendix A.4 of the book [Ber22a] and Prop.\ 2.3.1 of the book [Ber22b]. 
\item[(e)] The region of stability, i.e., the set of $\tl J$ for which $\tl \m$ is stable in the sense that 
$J_{\tl\m}(x)<\infty$ for all $x\in X,$
  is closely connected to 
the region of convergence of Newton's method. 
\item[(f)] The region of  stability is also enlarged by increasing the length of the rollout horizon, as long as the base policy that is used for rollout is stable. 
\item[(g)] Rollout with a stable policy $\m$ (i.e., $\tl J=J_\m$) guarantees that the lookahead policy $\tl \m$ is also stable, regardless of the length $\ell$ of lookahead.

\end{itemize}

In the next section, we will illustrate the preceding points through the use of a simple one-dimensional linear-quadratic problem, for which the Bellman equation can be defined through a one-dimensional Riccati equation. 
We note, however, that all the insights obtained through the Riccati equation survive intact to far more general problems, involving abstract Bellman equations where cost functions are defined over an arbitrary state space.\footnote{In this more general setting, the Bellman equation does not have the differentiability properties required to define  the classical form of Newton's method. However, Newton's method has been extended to nondifferentiable operator equations through the work of many researchers starting in the late 70s, and in a form that is perfectly adequate to support theoretically the DP/RL/MPC setting; see Josephy [Jos79], Robinson [Rob80], [Rob88], [Rob11], Kojima and Shindo [KoS86], Kummer [Kum88], [Kum00], Pang [Pan90], Qi and Sun [Qi93], [QiS93], Facchinei and Pang [FaP03], Ito and  Kunisch [ItK03], Bolte, Daniilidis, and Lewis [BDL09]. A  convergence analysis of the nondifferentiable form of Newton's method, together with a discussion of superlinear performance bounds that relate to MPC, is given in Appendix A of the book [Ber22a].}

In Section 3, we will briefly discuss stochastic extensions, where the system equation involves stochastic disturbances $w_k$:
$$x_{k+1}=f(x_k,u_k,w_k),\qquad k=0,1,\ldots.$$
The primary difficulty with stochastic problems is the increase of the computation required for both off-line training and on-line play, which may now involve Monte-Carlo simulation of $w_k$. This computation can be effectively mitigated with the use of {\it certainty equivalence\/}, i.e., by replacing the stochastic disturbances $w_k$ with typical values $\ol w_k$ (such as for example the expected values $E\{w_k\}$). However, it is essential that when performing the $\ell$-step lookahead minimization, {\it we use certainty equivalence only for the time steps $k+1,\ldots,{k+\ell-1}$, after the first step\/}. This is necessary in order to maintain the Newton step character of the on-line play process.  

In Section 4, we will comment on connections of the MPC/AlphaZero framework with adaptive control. An additional benefit of on-line policy generation by approximation in value space, not observed in the context of games (which have stable rules and environment), is that it works well with changing problem parameters and on-line replanning. Mathematically, what happens is that the Bellman equation is perturbed due to the parameter changes, but approximation in value space still operates as a Newton step. An essential requirement within this context is that a system model is estimated on-line through some identification method, and is used during the one-step or multistep lookahead minimization process, similar to what is done in indirect adaptive control. Within this context, we propose a simplified/faster version of indirect adaptive control, which uses rollout in place of policy reoptimization.


\section{Off-Line Training and On-Line Play Synergy Through Newton's Method}

We will now aim to understand the character of approximation in value space as it relates to the Bellman equation, and to the principal algorithms for its solution. To this end we will focus on  the one-dimensional  version of the classical linear-quadratic problem, where the system has the form
$$x_{k+1}=ax_k+bu_k.$$
Here the state $x_k$ and the control $u_k$ are scalars, and the coefficients $a$ and $b$ are also scalars, with $b\ne0$. The cost function has the form
$$\sum_{k=0}^\infty (qx_k^2+ru_k^2),$$
where $q$ and $r$ are positive scalars, and we assume no discounting ($\a=1$). 

This one-dimensional case admits a convenient and visually insightful analysis of the algorithmic issues that are central for our purposes. However, the analysis fully generalizes to multidimensional linear-quadratic problems. It also extends to general DP problems, including those involving arbitrary state and control spaces, stochastic or set membership uncertainty, as well as multiplicative/risk-sensitive cost functions. At this level of generality, the analysis requires a more demanding mathematical treatment that is based on the machinery of abstract DP; see the books [Ber20], [Ber22b].


\subsection{The Riccati Equation}

\begin{figure}
\begin{center}
\includegraphics[width=8.0cm]{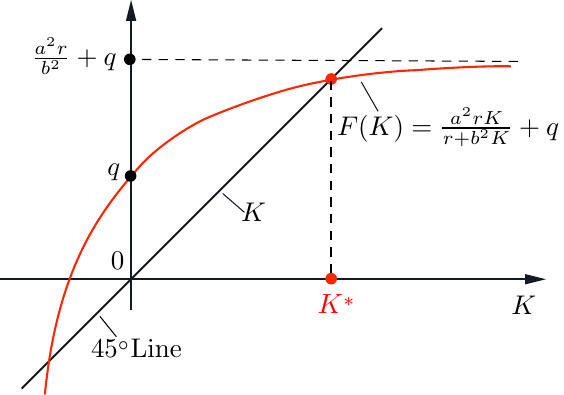}    
\caption{Graphical solution of the Riccati equation. The optimal cost function is $J^*(x)=K^*x^2$. The scalar $K^*$ solves the fixed point equation 
$K=F(K)$. It can be found graphically as the positive value of $K$ that corresponds to the point where the graphs of the functions $K$ and $F(K)$ meet. A similar interpretation can be given for the solution of the general Bellman equation, which however cannot be visually depicted for problems involving more than one or two states; see the books [Ber20], [Ber22a], and [Ber22b].
}
\label{figriccati}
\end{center}
\end{figure}

\pn Let us summarize the main analytical and computational results that we will need (all of these are well known and can be found in many sources, including nearly all textbooks on MPC and optimal control). The optimal cost function is quadratic of the form 
$$J^*(x)=K^*x^2,$$
where the scalar $K^*$ is the unique positive solution of  Riccati equation
$$K=F(K)={a^2 rK\over r+b^2K}+q.$$
Moreover,  the optimal policy is linear of the form
$$\m^*(x)=L^*x,$$
where $L^*$ is the scalar given by
$$L^*=-{abK^*\over r+b^2K^*}.$$ 
The Riccati equation is simply the Bellman equation  restricted to quadratic functions $J(x)=Kx^2$ with $K\ge0$. Both  the Riccati and the Bellman equations can be viewed as fixed point equations, and can be graphically interpreted and solved graphically as indicated in Fig.\ \ref{figriccati}.

\begin{figure}
\begin{center}
\includegraphics[width=8.4cm]{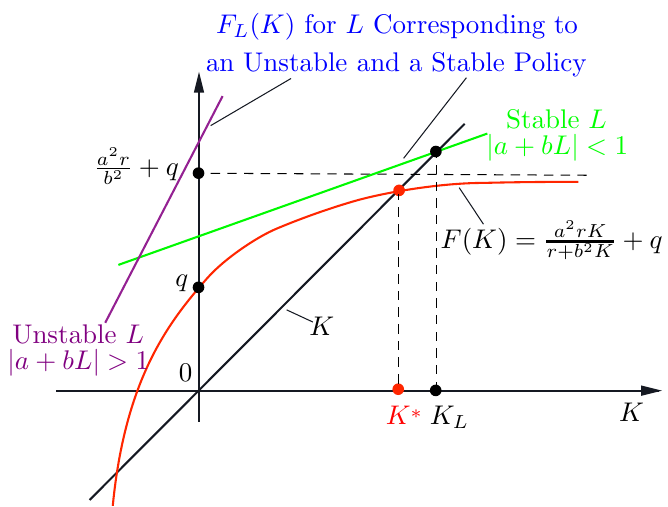}    
\caption{Graphical solution of the Riccati equation for a  linear policy $\m(x)=Lx$. When $\m$ is stable, its cost function is $J_\m(x)=K_Lx^2,$ where $K_L$ corresponds to the point where the graphs of the functions $K$ and $F_L(K)$ meet.}
\label{figriccatipol}
\end{center}
\end{figure}

We can also characterize graphically the cost function of a policy $\m$ that is linear of the form
$\m(x)=Lx$, and is also stable, in the sense that the scalar $L$ satisfies $|a+bL|<1$,
so that the corresponding closed-loop system
$$x_{k+1}=(a+bL)x_k$$
 is stable. Its cost function has the form
 $$J_\m(x)=K_Lx^2,$$
 where $K_L$ solves the equation\footnote{Sometimes this equation is called the ``Lyapunov equation" in the control theory literature. In this paper, we will refer to it as the ``Riccati equation for linear policies."}
 $$K=F_L(K)=(a+bL)^2K+q+rL^2.$$
 The graphical solution of this equation is illustrated in Fig.\ 
\ref{figriccatipol}. The function $F_L(K)$ is linear, with slope $(a+bL)^2$ that is strictly less than 1. In particular, $K_L$ corresponds to the point where the graphs of the functions $K$ and $F_L(K)$ meet.

If $\m(x)=Lx$  is unstable, in the sense that the scalar $L$ satisfies $|a+bL|>1$, then its cost function is given by $J_\m(x)=\infty$ for all $x\ne0$ and $J_\m(0)=0$. In this case the graphs of the functions $K$ and $F_L(K)$ meet at a negative value of $K$, which has no meaning in the context of the linear-quadratic problem. 

\subsection{Iterative Solution by Value and Policy Iteration}

\begin{figure}
\begin{center}
\includegraphics[width=8.0cm]{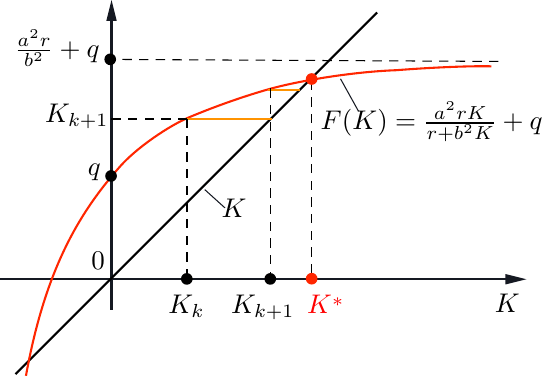}    
\caption{Graphical illustration of VI. It has the form $K_{k+1}=F(K_k)$, where $F$ is the Riccati operator,
$$F(K)={a^2 rK \over r+b^2 K }+q.$$ 
 The algorithm converges to $K^*$ starting from any $K_0\ge0$.}
\label{figriccativi}
\end{center}
\end{figure}

The classical DP algorithm of Value Iteration (VI for short) produces a sequence of cost functions $\{J_k\}$ by applying the Bellman equation operator repeatedly, starting from an initial nonnegative function $J_0$. For our linear-quadratic problem it takes the form
$$J_{k+1}(x)=\min_{u\in\Re}\big\{qx^2+ru^2+J_k(ax+bu)\big\}.$$
When $J_0$ is quadratic of  the form $J_0(x)=K_0x^2$ with $K_0\ge0$, it can be seen that the VI iterates $J_{k}$ are also quadratic of the form
$J_{k}(x)=K_{k}x^2$, where
$$K_k=F(K_{k-1}).$$
Then the VI algorithm becomes a fixed point iteration that uses the Riccati operator $F$. The algorithm is illustrated in Fig. \ref{figriccativi}. As can be seen from the figure, when starting from any $K_0\ge0$, the algorithm generates a sequence $\{K_k\}$ of nonnegative scalars that converges to $K^*$.

Another major algorithm is Policy Iteration (PI for short). It produces a sequence of stable policies $\{\m^k\}$, starting with some stable policy $\m^0$. Each policy has improved cost function over the preceding one, i.e., $J_{\m^{k+1}}(x)\le J_{\m^{k}}(x)$ for all $k$ and $x$, and the sequence of policies $\{\m^k\}$ converges to the optimal. Policy iteration is of major importance in RL, since most of the successful algorithmic RL schemes use explicitly or implicitly some form of approximate PI. We will discuss PI and its relation with rollout later, and we will provide visual interpretations based on their connection with Newton's method.

\subsection{Visualizing Approximation in Value Space}

The use of Riccati equations allows insightful visualization of approximation in value space. This visualization, although specialized to linear-quadratic problems, is consistent with related visualizations for more general infinite horizon problems. In particular, in the books [Ber20] and [Ber22a], Bellman operators, which define the Bellman equations,  are used in place of Riccati operators, which define the Riccati equations.

\begin{figure}
\begin{center}
\includegraphics[width=8.4cm]{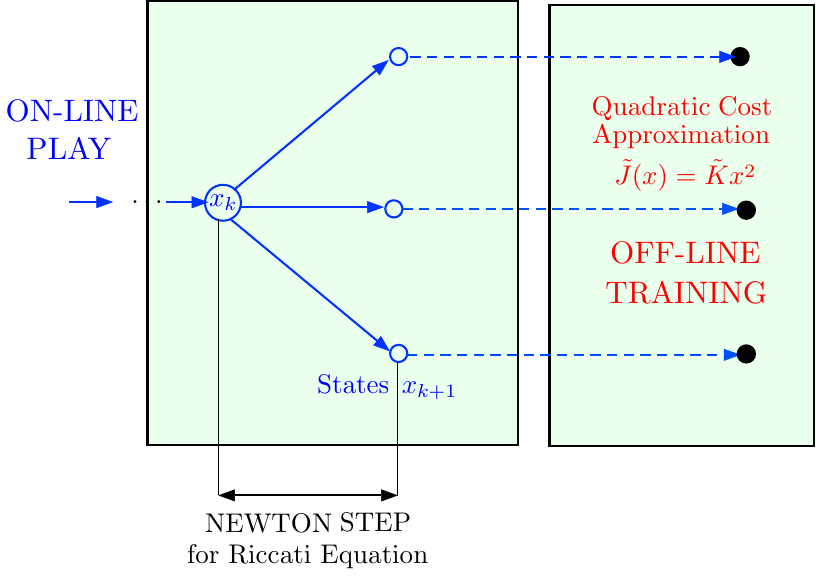}    
\caption{Illustration  of the interpretation of approximation in value space with one-step lookahead as a Newton step that maps $\skew5\tl J$ to the cost function $J_{\tl \m}$ of the one-step lookahead policy.}
\label{figonestepNewton}
\end{center}
\end{figure}

We will first show that approximation in value space with one-step lookahead can be viewed as a Newton step for solving the Riccati equation; see Fig.\ \ref{figonestepNewton}. In particular, let us consider a quadratic cost function approximation of the form $\tl J(x)=Kx^2$, where $K\ge0$. We will show that:
\begin{itemize}
\item[(a)] An iteration of Newton's method for solving the Riccati equation $K=F(K)$, starting from a value $\tl K$ yields the quadratic cost coefficient $K_{\tl L}$ of the cost function $J_{\tl \m}$ of the one-step lookahead policy ${\tl \m}$, which is linear of the form ${\tl \m}(x)=\tl L x$ and has cost function $J_{\tl \m}(x)=K_{\tl L}x^2$. 
\item[(b)]As a result of (a), the  quadratic cost coefficients $\tl K$  and  $K_{\tl L}$ satisfy the quadratic convergence relation
$${|K_{\tl L}-K^*|\over |\tl K-K^*|^2}<\infty.$$
\item[(c)]As a result of (b), for $\tl K$ within the region of convergence of Newton's method, the one-step lookahead policy cost function $J_{\tl \m}$  tends to be closer to $J^*$ than $\tl  J$, and for $\tl J$ close to $J^*$, the policy $\tl \m$ is very close to optimal.  
\end{itemize}
These facts admit a simple proof for the linear-quadratic case, but qualitatively hold in great generality, i.e., for arbitrary state and control spaces, for finite and infinite horizon problems, and in the presence of stochastic and set-membership uncertainty. The reason for this generality is the universal character of the corresponding mathematical proof arguments, which rely on the theory of abstract DP.

\begin{figure}
\begin{center}
\includegraphics[width=8.4cm]{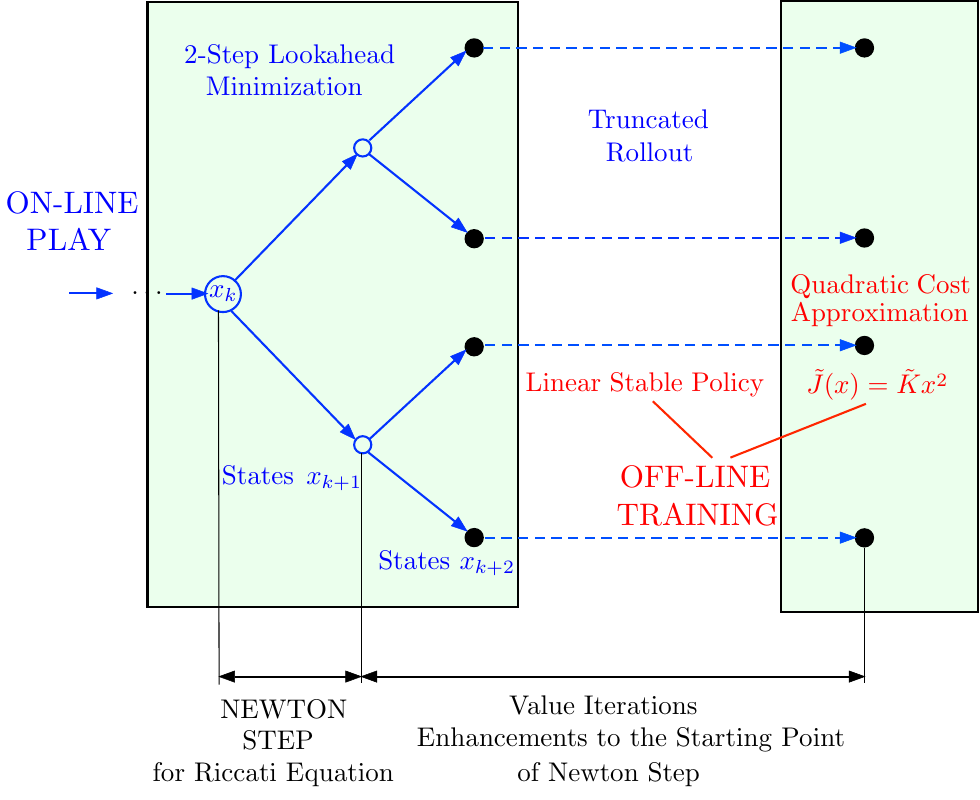}
\caption{Illustration  of the interpretation of approximation in value space with multistep lookahead and truncated rollout as a Newton step, which maps the result of multiple VI iterations starting with the terminal cost function approximation $\skew5\tl J$ to the cost function $J_{\tl \m}$ of the multistep lookahead policy.}
\label{figmultistepNewton}
\end{center}
\end{figure}

For the case of multistep lookahead minimization, which typically underlies the MPC architecture, we will also show that the Newton step property holds. Indeed, this property is enhanced, because {\it the region of convergence of Newton's method is enlarged by longer lookahead\/}, as we will argue graphically later. The extension of the Newton step interpretation is not surprising because, as noted earlier, we may view $\ell$-step lookahead as a one-step lookahead where the cost function approximation is the optimal cost function of an $(\ell-1)$-stage DP problem with a terminal cost $\tl J(x_{k+\ell})$ on the state $x_{k+\ell}$ obtained after $\ell-1$ stages; see Fig.\ \ref{figmultistepNewton}.

\begin{figure}
\begin{center}
\includegraphics[width=8.4cm]{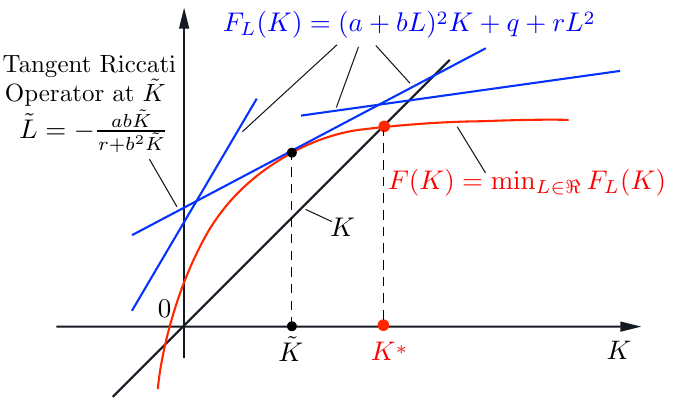}    
\caption{Illustration  of how the graph of the Riccati operator $F$ can be obtained as the lower envelope of the linear operators 
$$F_L(K)=(a+bL)^2K+q+bL,$$
 as $L$ ranges over $\Re$, i.e.\
$F(K)=\min_{L\in \Re}F_L(K).$
Moreover, for any fixed $\tl K$, the scalar $\tl L$ that attains the minimum is given by 
$\tl L=-{ab{\tl K}\over r+b^2{\tl K}},$
and is such that the line corresponding to the graph of $F_{\tl L}$ is tangent to the graph of $F$ at $\tl K$, as shown in the figure.}
\label{figenvelope}
\end{center}
\end{figure}

Indeed,  let us first consider one-step lookahead minimization with any terminal cost function approximation of the form $\tl J(x)= Kx^2$, where $K\ge0$. The one-step lookahead control at state $x$, which we denote by $\tl \m(x)$, is obtained by minimizing the right side of Bellman's equation when $J(x)=Kx^2$:
$$\tl \m(x)\in\arg\min_{u\in\Re}\big\{qx^2+ru^2+K(ax+bu)^2\big\}.$$
We can break this minimization into a sequence of two minimizations as follows:
\begin{align*}
F(K)x^2&=\min_{L\in \Re}\min_{u=Lx}\big\{qx^2+ru^2+K(ax+bu)^2\big\}\\
&=\min_{L\in \Re}\big\{q+bL+ K(a+bL)^2\big\}x^2\\
&=\min_{L\in \Re}F_L(K)x^2,
\end{align*}
where the function $F_L(K)$ is the Riccati equation operator for the generic linear policy $\m(x)=Lx$.
Figure \ref{figenvelope} illustrates the two-step minimization of the preceding equation, and shows how the graph of {\it the Riccati operator $F$ can be obtained as the lower envelope of the linear operators $F_L$\/}, as $L$ ranges over the real numbers. 

\begin{figure}
\begin{center}
\includegraphics[width=8.0cm]{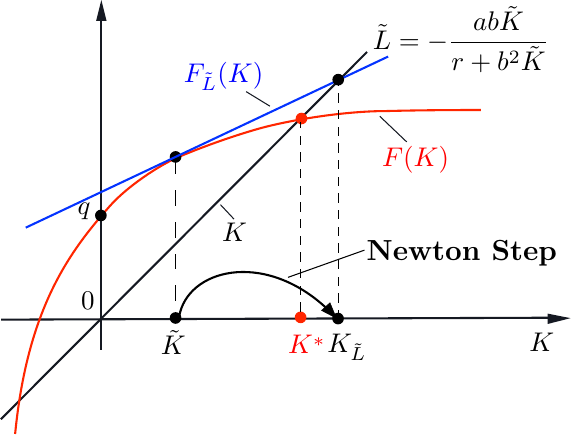}    
\caption{Illustration of approximation in value space with one-step lookahead. Given a terminal cost approximation $\skew 5\tl J=\skew 3\tl Kx^2$, we compute the corresponding linear policy $\tl \m(x)=\tl Lx$, where
 $$\tl L=-{ab\tl K\over r+b^2\tl K},$$
and the corresponding cost function $K_{\tl L}x^2$, using the Newton step shown.}
\label{figvalspacelqonestep}
\end{center}
\end{figure}

Let us now fix the terminal cost function approximation to some $\tl Kx^2$, where $\tl K\ge0$, and consider the corresponding one-step lookahead policy $\tl \m$. Figure \ref{figvalspacelqonestep} illustrates the corresponding linear cost function  
$F_{\tl L}$ of  $\tl \m$, and shows that its graph is a tangent line to the graph of $F$ at the point $\tl K$
(cf.\ Fig.\ \ref{figenvelope}). 

Thus the function $F_{\tl L}$ can be viewed as a linearization of $F$ at the point $\tl K$, and defines a linearized problem: to find a solution of the equation 
$$K=F_{\tl L}(K)=q+b\tl L^2+K(a+b\tl L)^2.$$
The important point now is that {\it the solution of this equation, denoted $K_{\tl L}$, is the same as the one obtained from a single iteration of Newton's method for solving the Riccati equation, starting from the point $\tl K$\/}. This is illustrated in Fig.\ \ref{figvalspacelqonestep}.

To elaborate, let us note that the classical form of Newton's method for solving a fixed point problem of the form $y=T(y)$, where $y$ is an  $n$-dimensional vector, operates as follows: At the current iterate $y_k$, we linearize $T$ and find the solution $y_{k+1}$ of the corresponding linear fixed point problem. Assuming $T$ is differentiable, the linearization is obtained by using a first order Taylor expansion:
$$y_{k+1}=T(y_k)+{\partial T(y_k)\over \partial y}(y_{k+1}-y_k),$$
where ${\partial T(y_k)/\partial y}$ is the $n\times n$ Jacobian matrix of $T$ evaluated at the vector $y_k$. For the linear quadratic problem, $T$ is equal to the Riccati operator $F$, and is differentiable. However, there are extensions of Newton's method that are based on solving a linearized system at the current iterate, but relax the differentiability requirement to piecewise differentiability, and/or component concavity (here the role of the Jacobian matrix is played by subgradient operators).  The quadratic or similarly fast superlinear convergence property is maintained in these extended forms of Newton's method; see  the monograph  [Ber22a] (Appendix A) and the paper [Ber22c], which provide a convergence analysis and discussion related to the DP/MPC context.

\begin{figure}
\begin{center}
\includegraphics[width=8.0cm]{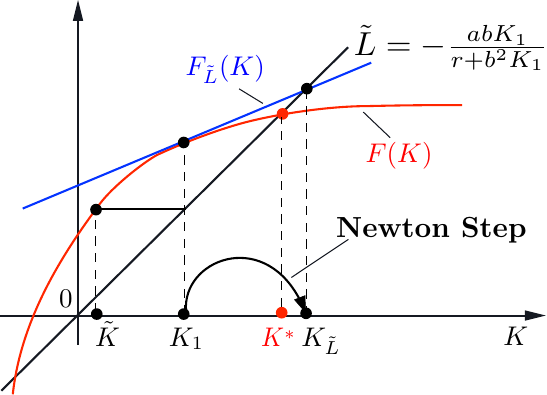}    
\caption{Illustration of approximation in value space with two-step lookahead. Starting with a terminal cost approximation $\skew 5\tl J=\skew 3\tl Kx^2$, we obtain $K_1$ using a single value iteration. We then compute the corresponding linear policy $\tl \m(x)=\tl Lx$, where
$$\tl L=-{abK_1\over r+b^2K_1}$$
and the corresponding cost function $K_{\tl L}x^2$, using the Newton step shown. The figure shows that for any $K\ge0$, the corresponding $\ell$-step lookahead policy will be stable for all $\ell$ larger than some threshold.}
\label{figvalspacelqmultistep}
\end{center}
\end{figure}

The preceding argument can be extended to $\ell$-step lookahead minimization to show that a similar Newton step interpretation is possible (Fig.\ \ref{figvalspacelqmultistep} depicts the case $\ell=2$). Indeed in this case, instead of linearizing $F$ at $\tl K$, we linearize at 
$$K_{\ell-1}=F^{\ell-1}(\tl K),$$
 i.e., at the result of $\ell-1$ successive applications of $F$ starting with $\tl K$. Each application of $F$ corresponds to a value iteration. Thus {\it the effective starting point for the Newton step is $F^{\ell-1}(\tl K)$\/}.

\begin{figure}
\begin{center}
\includegraphics[width=8.0cm]{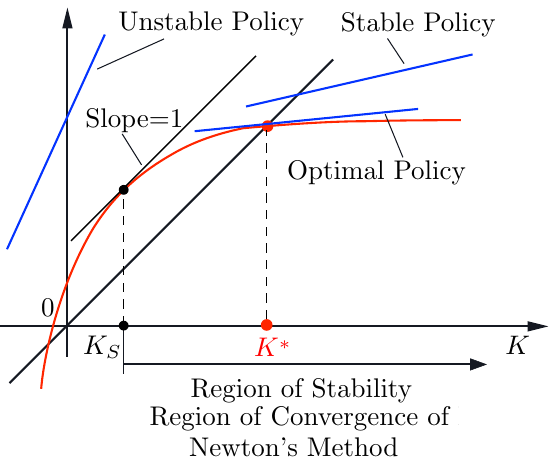}    
\caption{Illustration of the region of stability, i.e., the set of $K\ge0$ such that the one-step lookahead policy is stable. This is also the set of initial conditions for which Newton's method converges to $K^*$ asymptotically. 
}
\label{figstability}
\end{center}
\end{figure}

\subsection{Region of Stability of Approximation in Value Space}

It is  useful to define the {\it region of stability} of approximation in value space as the set of $K\ge0$ such that 
$$|a+bL_K|<1,$$ 
where $L_K$ is the linear coefficient of the one-step lookahead policy corresponding to $K$.
It can be seen that the region of stability is also closely related to {\it the region of convergence of Newton's method\/}: the set of  points $K$ starting from which Newton's method, applied to the Riccati equation $K=F(K)$, converges to $K^*$ asymptotically. 

Note that for our one-dimensional linear-quadratic problem, the region of stability is the interval $(K_S,\infty)$ that is characterized by the single point $K_S$ where $F$ has derivative equal to 1; see Fig.\ \ref{figstability}. 
For multidimensional problems, the region of stability may not be characterized as easily. Still, however, it is generally true that {\it the region of stability is enlarged as the length of the lookahead increases\/}. Moreover, substantial subsets of the region of stability may be conveniently obtained. Results of this type are known within the MPC framework under various conditions (see the papers by Mayne at al.\ [MRR00], Magni et al.\ [MDM01], and the MPC book [RMD17]). 

In this connection, it is interesting to note that with increased lookahead, the effective starting point 
$F^{\ell-1}(\tl K)$ is pushed more and more within the region of stability, and approaches $K^*$ as $\ell$ increases. In particular, it can be seen that {\it for any given $\tl K\ge0$, the corresponding $\ell$-step lookahead policy will be stable for all $\ell$ larger than some threshold\/}; see Fig.\ \ref{figvalspacelqmultistep}. The book [Ber22a], Section 3.3, contains a broader discussion of the region of stability and the role of multistep lookahead in enlarging it. 

\subsection{Rollout and Policy Iteration}

\begin{figure}
\begin{center}
\includegraphics[width=8.4cm]{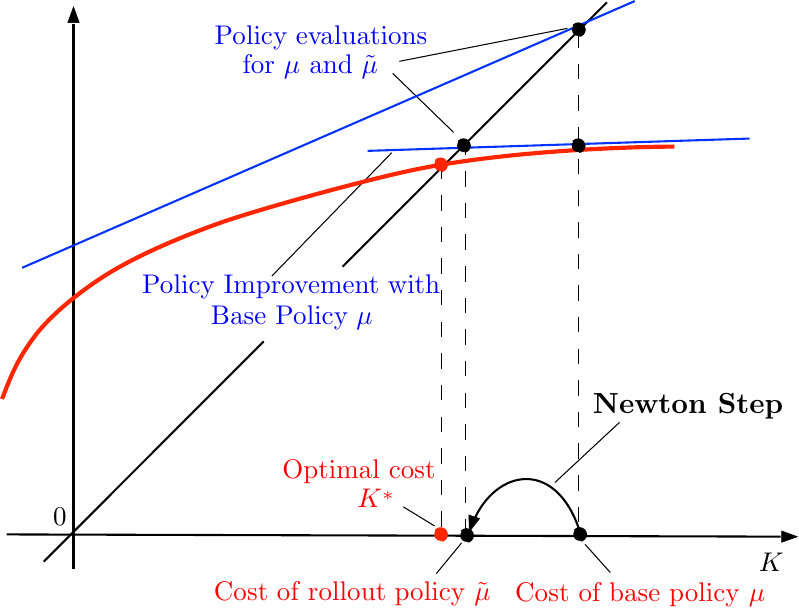}    
\caption{Illustration of the rollout algorithm. Starting from a linear stable base policy $\m$, it generates a stable rollout policy $\tl \m$. The quadratic cost  coefficient of $\tl \m$ is obtained from the quadratic cost  coefficient of $\m$ with a Newton step for solving the Riccati equation.}
\label{figrolloutlqonestep}
\end{center}
\end{figure}

Let us return to the linear quadratic problem and the rollout algorithm starting from a  stable linear base policy $\m$. It obtains the rollout policy $\tl \m$ by using a policy improvement operation, which by definition, yields the one-step lookahead policy that corresponds to terminal cost approximation $J_\m$. 
Figure \ref{figrolloutlqonestep} illustrates the rollout algorithm. It can be seen from the figure that the rollout policy is in fact an improved policy, in the sense that $J_{\tl \m}(x)\le J_\m(x)$ for all $x$, something that is true in general (not just for linear-quadratic problems). Among others, this implies that the rollout policy is stable. 

Since the rollout policy is a one-step lookahead policy, it  can also be described using the formulas that we developed earlier in this section. In particular, let the base policy have the form
$$\m^0(x)=L_0x,$$ 
where $L_0$ is a scalar. We require that $\m^0$ is stable, i.e., $|a+bL_0|<1$. From our earlier calculations, we have that the cost function of $\m^0$ is
$$J_{\m^0}(x)=K_0x^2,$$
where
$$K_0={q+rL_0^2\over 1-(a+bL_0)^2}.$$
Moreover, the rollout policy $\m^1$ has the form
$\m^1(x)=L_1x,$
where
$$L_1=-{abK_0\over r+b^2K_0}.$$

We can similarly describe the  policy iteration (PI) algorithm. It is simply the repeated application of rollout, and generates a sequence of stable linear policies $\{\m^k\}$. By replicating our earlier calculations, we see that these policies have the form
 $$\m^k(x)=L_k x,\qquad k=0,1,\ldots,$$
 where $L_k$ is generated by the iteration
 $$L_{k+1}=-{a bK_k\over r+b^2K_k},$$
 with $K_k$ given by 
$$K_k={q+rL_k^2\over 1-(a+bL_k)^2}.$$

The corresponding cost functions have the form 
$$J_{\m^k}(x)=K_kx^2.$$
A favorable characteristic that enhances the performance of rollout and PI is that the graph of $F(K)$ is relatively ``flat" for $K>K^*$. This is due to the concavity of the Riccati operator. As a result, the cost improvement due to the Newton step is even more pronounced, and is relatively insensitive to the choice of base policy. This feature generalizes to multidimensional problems with or without constraints; see the computational study [LKL23]. 

Part of the classical linear-quadratic theory is that $J_{\m^k}$ converges to the optimal cost function $J^*$, while the generated sequence of linear policies $\{\m^k\}$, where $\m^k(x)=L_kx$, converges to the optimal policy, assuming that the initial policy is linear and stable. The convergence rate of the sequence $\{K_k\}$ is quadratic, as is typical of Newton's method. This result was proved by Kleinman [Kle68] for the continuous-time multidimensional version of the linear quadratic problem, and it was extended later to more general problems. In particular, the corresponding discrete-time result was given by Hewer [Hew71], and followup analysis, which relates to policy iteration with approximations, was given by Feitzinger, Hylla, and Sachs [FHS09], and Hylla [Hyl11]. Kleinman gives credit to Bellman and Kalaba [BeK65] for the one-dimensional version of his results.
Applications of approximate PI in the context of MPC have been discussed in Rosolia and Borrelli [RoB18], and Li et al.\ [LJM21], among others.

It is important to note that rollout, like policy iteration, can be applied universally, well beyond the linear-quadratic/MPC context that we have discussed here. In fact, the main idea of rollout algorithms, obtaining an improved policy starting from some other suboptimal policy, has appeared in several DP contexts, including games; see e.g., Abramson [Abr90], and Tesauro and Galperin [TeG96]. The adaptation of rollout to discrete deterministic optimization problems and the principal results relating to cost improvement were given in the paper by Bertsekas,
Tsitsiklis, and Wu [BTW97], and were also discussed in the neuro-dynamic programming book [BeT96]. Rollout algorithms for stochastic problems were further formalized in the papers by   Bertsekas [Ber97], and Bertsekas and Casta\~ non [BeC99]. Extensions to constrained rollout were given in the author's papers [Ber05a], [Ber05b]. Rollout algorithms were also proposed in nontruncated form within the MPC framework; see De Nicolao, Magni, and Scattolini [DMS98], [MaS04], and followup works.

A noteworthy extension, highly relevant to MPC as well as other contexts, is {\it multiagent rollout\/}, which deals successfully with the acute computational difficulties arising from the large (Cartesian product) control spaces that are typical of multiagent problems. The author's book [Ber20] and paper [Ber21a] discuss this research, and give references to supportive computational studies in multi-robot and vehicle routing problems with imperfect state information, among others; see [BKB20], [GPG22], and [WGP23].

Finally, we note that the author's books [Ber20], [Ber22a],  [Ber23] provide extensive references to the journal literature, which includes a large number of computational studies.
These studies discuss variants and problem-specific adaptations of rollout algorithms and consistently report  favorable computational experience. The size of the cost improvement over the base policy is often impressive, evidently owing to the fast convergence rate of Newton's method that underlies rollout.

\vspace{-0.5pc}
\subsection{Truncated Rollout}

\begin{figure}
\begin{center}
\includegraphics[width=8.4cm]{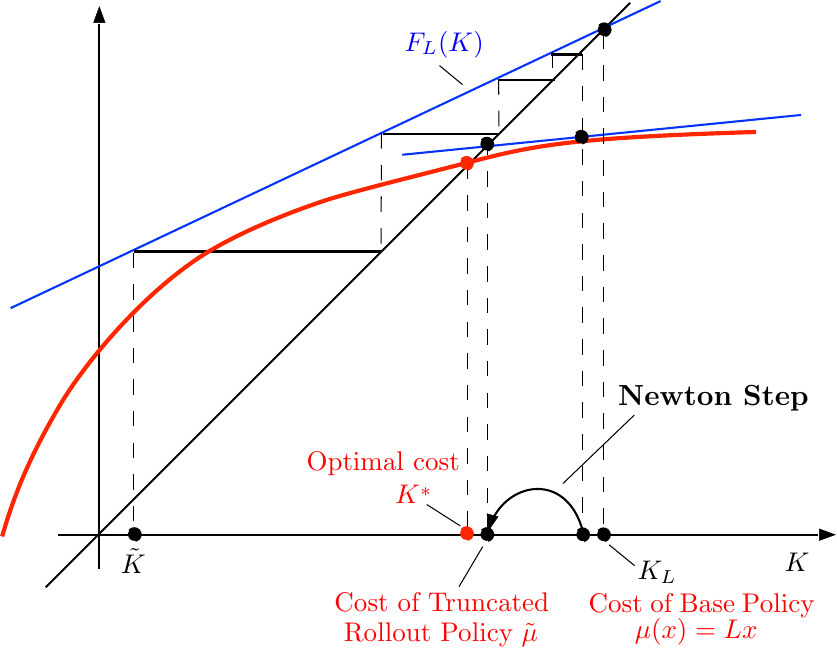}    
\caption{Illustration of the $m$-step truncated rollout algorithm with one-step lookahead. Starting with a linear stable base policy $\m(x)=Lx$, it generates a rollout policy $\tl \m$. The quadratic cost  coefficient of $\tl \m$ is obtained with a Newton step, after approximating of the quadratic cost  coefficient $K_L$ of $\m$ with $m=4$ value iterations that start from $\tl K$. Compare with the nontruncated rollout Fig.\ \ref{figrolloutlqonestep}.}
\label{figstabilitylqmultistep}
\end{center}
\end{figure}

An $m$-step truncated rollout scheme with a stable linear base policy $\m(x)=Lx$,  one-step lookahead minimization, and terminal cost approximation $\tl J(x)=\tl Kx^2$  is obtained by starting at $\tl K$, executing $m$ VI steps using $\m$, followed by a one-step lookahead minimization/Newton step. It is visually interpreted as in Fig.\ \ref{figstabilitylqmultistep}, where $m=4$. 

Thus the difference with (nontruncated) rollout is that we use $m$ VI steps starting from $\tl K$ to approximate the cost function $K_Lx^2$ of the base policy. Truncated rollout makes little sense in linear-quadratic problems where $K_L$ can be easily computed by solving the Riccati equation. However, it is useful in more general problem settings, as it may save significantly in computation, relative to obtaining exactly $J_{\m}$ (which requires an infinite number of VI steps).

Some interesting points regarding truncated rollout schemes are the following:
\begin{itemize}
\item[(a)]  Lookahead by truncated rollout may be an economic substitute for lookahead by minimization, in the sense that it may achieve a similar performance at significantly reduced computational cost; see e.g., [LiB24]. 
\item[(b)]  Lookahead by $m$-step truncated rollout with a stable policy has an increasingly beneficial effect on the stability properties of the lookahead policy, as $m$ increases.
\end{itemize}
These statements are difficult to establish analytically in some generality. However, they can be intuitively understood in the context with our one-dimensional linear quadratic problem, using geometrical  constructions like the one of Fig.\ \ref{figstabilitylqmultistep}. They are also consistent with the results of computational experimentation. We refer to the monograph [Ber22a] for further discussion.

\subsection{Double Rollout}

We noted that rollout with a base policy $\m$ amounts to a single policy iteration starting with $\m$, to produce the (improved) rollout policy $\tl \m$. The process can now be continued to apply a second policy iteration. This results in a {\it double rollout} policy, i.e., a second rollout policy that uses the first rollout policy $\tl \m$ as a base policy. For deterministic problems, the needed rollout policy costs can be computed recursively on-line, with computation that may be tractable, thanks to rollout truncation or special simplifications that take advantage of the deterministic character of the problem. Parallel computation, for which rollout is  well suited, can also be very helpful in this respect.

Triple and higher order rollout, which amount to multiple successive policy iterations, are possible. However, the on-line computational costs quickly become overwhelming, despite the potential use of truncation and other simplifications, or parallel computation.

For further discussion of double rollout, see Section 2.3.5 of the  book [Ber20] and Section 6.5 of the book [Ber22], and for computational experimentation results, see the recent paper by Li and Bertsekas [LiB24], which deals with special inference contexts in hidden Markov models.  Policy iteration/double rollout has also been discussed by Yan et al.\ [YDR04] in the context of the game of solitaire, and by Silver and Barreto  [SiB22] in the context of a broader class of search methods.

\subsection{Double Newton Step - Rollout on Top of Approximation in Value Space}

\begin{figure}
\begin{center}
\includegraphics[width=8.4cm]{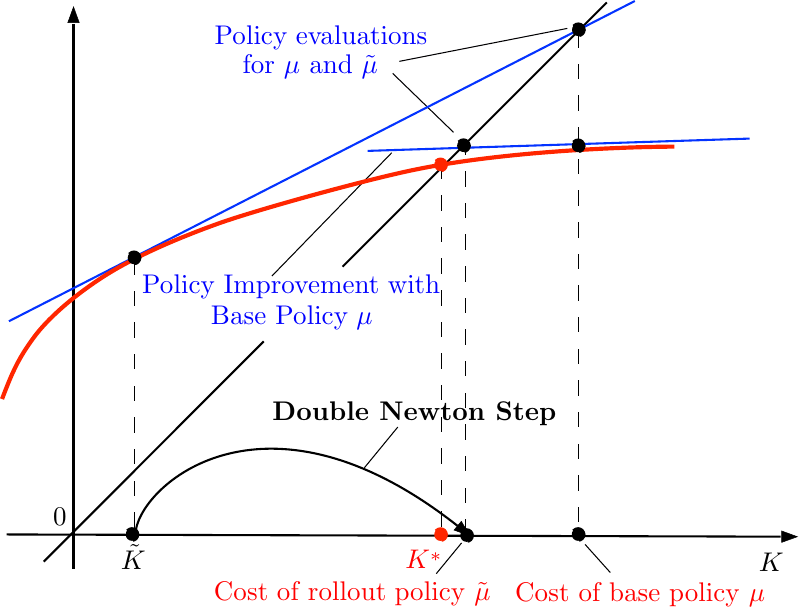}    
\caption{Illustration of a double Newton step. Starting from a quadratic cost coefficient $\tl K$ that defines the  policy $\m$, it uses $\m$ as a base policy to implement a rollout policy $\tl \m$.}
\label{figdoublerollout}
\end{center}
\end{figure}

Given a quadratic cost coefficient $\tl K$ that defines the  policy $\m(x)=L_{\tl K}x$, it is natural and convenient to consider rollout that uses $\m$ as a base policy. This can be viewed as rollout that is built on top of approximation in value space. We  call this algorithm {\it double Newton step\/}, because it consists of two Newton steps: a first step that maps $\tl Kx^2$ to $J_{\m}(x)$ and a second step that maps $J_{\m}(x)$ to the cost function $J_{\tl\m}(x)$ of the rollout policy $\tl \m$ that is produced when the base policy is $\m$; see Fig.\ \ref{figdoublerollout}. 

The double Newton step is much more powerful than the algorithm that performs approximation in value space with two-step lookahead starting from $\tl K$. In particular, both algorithms involve multiple steps for solving the Riccati equation starting from  $\tl K$. However, the former algorithm amounts to a Newton step followed by a Newton step, while the later algorithm amounts to a value iteration followed by a Newton step (cf.\ Fig.\ \ref{figvalspacelqmultistep}). For this statement to be correct, $\tl K$ should lie within the region of stability. Such $\tl K$  may obtained by using multiple value iterations, as in the case where a multistep lookahead minimization is performed, i.e.\  $\ell>1$ (cf.\ Fig.\ \ref{figvalspacelqmultistep}).

Note that it is also possible to consider variants of rollout on top of approximation in value space, such as truncated and simplified versions. An important example of the truncated version is the TD-Gammon architecture, where the terminal cost function approximation is constructed off-line by using a neural network.

\subsection{The Importance of the First Step of  Lookahead}

\pn The Newton step interpretation of approximation in value space leads to an important insight into the special character of the initial step in $\ell$-step lookahead implementations. In particular, {\it it is only the first step that acts as the Newton step\/}, and needs to be implemented with precision; cf.\ Fig.\ \ref{figmultistepNewton}. The subsequent $\ell-1$ steps consist of a sequence of value iterations  with starting point $\a^\ell \tl J$, and only serve to enhance the quality of the starting point of the Newton step. As a result, {\it their precise implementation is not critical\/}; this is a major point in the narrative of the author's book [Ber22a].

This idea suggests that we can simplify (within reason) the lookahead steps after the first with small (if any) performance loss for the multistep lookahead policy. An important example of such a simplification is the use of certainty equivalence, which will be discussed in the next section. 
Other possibilities include  various ways of ``pruning"  the lookahead tree; see [Ber23], Section 2.4.  On the other hand, pruning the lookahead tree at the first stage of lookahead, as is often done in Monte Carlo Tree Search, can have a serious detrimental effect on the quality of the MPC policy.

In practical terms, simplifications after the first step of the multistep lookahead can save a lot of on-line computation, which can be fruitfully invested in extending the length of the lookahead. This insight is supported by substantial computational experimentation, starting with the paper by Bertsekas and Casta\~non [BeC98], which verified the beneficial effect of certainty equivalence (after the first step) as a rollout simplification device for stochastic problems.  On the other hand, implementing imprecisely the minimization of the first step can adversely impact the performance of  the multistep lookahead policy. This point is often missed in the design of approximate lookahead minimization schemes, such as Monte Carlo Tree Search.

\subsection{Newton Step Interpretation of Approximation in Value Space in General Infinite Horizon Problems}

\pn The interpretation of approximation in value space as a Newton step, and related notions of stability that we have discussed in this section admit a broad generalization. The key fact in this respect is that our DP problem formulation allows arbitrary state and control spaces, both discrete and continuous, and can be extended even further to general abstract models with a DP structure; see the abstract DP book [Ber22b]. 

Within this more general context, the Riccati operator is replaced by an abstract Bellman operator and the quadratic terminal cost function $\tl Kx^2$ is replaced by a general cost function $\tl J$. Valuable insight can be obtained from graphical interpretations of the Bellman equation, the VI and PI algorithms, one-step and multistep approximation in value space, the region of stability, and exceptional behavior; see the book [Ber22a], and  Section 1.6.7 of the book [Ber23] for a discussion of the MPC context.
Naturally, the graphical interpretations and visualizations are limited to one dimension. However, the visualizations provide insight, and motivate conjectures and mathematical proof analysis, much of which is given in the books [Ber20] and [Ber22a].

\subsection{How Approximation in Value Space Can Fail and What to Do About It}

\pn Practice has shown that MPC is a reliable methodology that can be made to work, assuming (as we have in this section) that a system model is available in either analytical form or in simulator form, and that this model is not changing over time. Still, however, even under these favorable circumstances, failure is possible, in the sense that the $\ell$-step lookahead MPC policy is performing poorly. Typically the reason for failure is that the terminal cost approximation $\tl J$ lies outside the region of convergence of the Newton step. This region depends on $\ell$ (see the discussion near  the end of Section 2.4), as well as the truncated rollout scheme, which effectively modifies  the starting point of the Newton step (see the discussion of Section 2.6).\footnote{In the case of the linear quadratic problem with terminal cost approximation $\tl J(x)=\tl Kx^2$, $\ell$-step lookahead minimization, and $m$-step truncated rollout with stable policy $\m(x)=Lx$, the region of stability is the set of all $\tl K$ such that $F^{\ell-1}\big(F_L^m(\tl K)\big)$ belongs to the set of $K$ such that $|a+bL_K|<1$; see Section 2.4 and Fig.\ \ref{figstability}.}

For an example of broad interest, let us assume that $\tl J$ is obtained by training with data a neural network (e.g., as in AlphaZero and TD-Gammon). Let us also focus  on the case of one-step lookahead with no truncated rollout. In this case there are three components that determine the approximation error $\tl J-J^*$:
\begin{itemize}
\item[(a)] The {\it power of the neural network architecture\/}, which roughly speaking is a measure of the error that would be obtained if infinite data were available and  used optimally to obtain $\tl J$ by training the given  neural network.
\item[(b)] The additional {\it error degradation due to limited availability of training data\/}.
\item[(c)]  The additional {\it error degradation due to imperfections in the training methodology\/}.
\end{itemize}
Thus {\it if the architecture is not powerful enough to bring $\tl J-J^*$ within the region of convergence of Newton's method, approximation in value space with one-step lookahead will likely fail, no matter how much data is collected and how effective the associated training method is\/}. 

In this case, there are two potential practical remedies: 
\begin{itemize}
\item[(1)] Use a more powerful architecture/neural network for representing $\tl J$, so it can be brought closer to $J^*$.
\item[(2)] Extend the combined length of the lookahead minimization and truncated rollout in order to bring the effective value of $\tl J$ within the region of convergence of Newton's method.
\end{itemize}

The first remedy typically requires a deep neural network or transformer, which uses more weights and requires more expensive training.\footnote{For a recent example of implementation of a grandmaster-level chess program with {\it one-step lookahead} and a huge-size (270M parameters) neural network position evaluator, see  Ruoss et al.\ [RDM24].} The second remedy requires longer on-line computation and/or simulation, which may run to difficulties because of some practical implementation limits. Parallel computation and sophisticated multistep lookahead methods may help to mitigate these requirements (see the corresponding discussions in the books [Ber22a] and [Ber23]).

\section{The Treatment of Stochastic Uncertainty Through Certainty Equivalence}

The main ideas of our framework extend to the case of a stochastic system of the form\footnote{In this section we restrict ourselves to stochastic uncertainty. For a parallel development relating to set-membership uncertainty and a minimax viewpoint, we refer to the books [Ber22a], Section 6.8,  [Ber22b], Chapter 5, and [Ber23], Section 2.12. The paper [Ber21b] addresses the challenging issue of convergence of Newton's method, applied to the Bellman equation of sequential zero-sum Markov games and minimax control problems. The zero-sum game structure differs in a fundamental way from its one-player optimization counterpart: its Bellman equation mapping need not be concave, and this complicates the convergence properties of Newton's method. The paper [Ber21b]  proposes new PI algorithms for discounted infinite horizon Markov games and minimax control, which are globally convergent, admit distributed asynchronous implementations, and lend themselves to the use of rollout and other RL methods.}
$$x_{k+1}=f(x_k,u_k, w_k),\qquad k=0,1,\ldots,$$
where $w_k$ is random with given probability distribution that depends only on the current state $x_k$ and control $u_k$, and not on earlier states and controls. The cost per stage also depends on $w_k$ and is $g(x_k,u_k,w_k)$.

The cost function of $\m$, starting from an initial state $x_0$ is
 $$J_\m(x_0) =
\lim_{N\tends\infty}E\left\{\sum_{k=0}^{N-1}
\a^kg\big(x_k,\mu(x_k),w_k\big)\right\},$$
where $E\{\cdot\}$ denotes expected value. The optimal cost function
$$J^*(x)=\min_{\m\in{\cal M}}J_\m(x),$$
again satisfies the Bellman equation, which now takes the form
$$J^*(x)=\min_{u\in U(x)}E\left\{g(x,u,w)+\a J^*\big(f(x,u,w)\big)\right\},\quad x\in X.$$ 
Furthermore, if $\m^*(x)$ attains the minimum above for all $x$, then $\m^*$ is an optimal policy. 

Similar to the deterministic case, approximation in value space with one-step lookahead replaces $J^*$ with an approximating function  $\tl J$, and obtains a suboptimal policy $\tl \m$ with the minimization 
$$\tl \m(x)\in\arg\min_{u\in U(x)}E\Big\{g(x,u,w)+\a \tl J\big(f(x,u,w)\big)\Big\},\quad x\in X.$$
It is also possible to use $\ell$-step lookahead, with the aim to improve the performance of the policy obtained through approximation in value space. This, however, can be computationally expensive, because the lookahead graph expands fast as $\ell$ increases, due to the stochastic character of the problem. Using {\it certainty equivalence} (CE for short) is an important approximation approach for dealing with this difficulty, as it reduces the search space of the $\ell$-step lookahead minimization. Moreover, CE mitigates the excessive simulation because it reduces the stochastic variance of the lookahead calculations at each stage. 

In the pure but somewhat flawed version of the CE approach, when solving the $\ell$-step lookahead minimization problem, we simply replace {\it all} of the uncertain quantities $w_k,w_{k+1},\ldots,w_{k+\ell-1},\ldots,w_{N-1}$ by some fixed nominal  values, thus making that problem fully deterministic. Unfortunately, this affects significantly the character of the approximation: when $w_k$ is replaced by a deterministic quantity, the  Newton step interpretation of the underlying approximation in value space scheme is lost to a great extent.

Still, we may largely correct this difficulty, while retaining substantial simplification, by using CE  {\it after the first  stage} of the $\ell$-step lookahead. We can do this with a CE scheme whereby at state $x_k$, we replace only  the uncertain  quantities $w_{k+1},\ldots,w_{N-1}$ by deterministic values, while we treat the first, i.e.,  $w_k$, as a stochastic quantity.\footnote{Variants of the CE approach, based on less drastic simplifications of the probability distributions of the uncertain  quantities, which involve multiple representative scenarios, are given in the author's books [Ber17a], Section 6.2.2, and [Ber19a], Section 2.3.2. Related ideas have also been suggested  in  MPC contexts; see e.g., Lucia, Finkler, and Engell [LFE13].}

This type of CE approach, first proposed and tested in the paper by Bertsekas and Casta\~non [BeC99], has an important property: {\it it maintains the Newton step character of the approximation in value space scheme\/}. 
In particular, the cost function $J_{\tl\m}$ of the $\ell$-step lookahead policy $\tl \m$ is generated by a Newton step, applied to the function obtained by the last $\ell-1$ minimization steps (modified by CE, and applied to the terminal cost function approximation); see the monograph [Ber20] and  Sections 1.6.7, 2.7.2, 2.8.3, of the textbook [Ber23] for a discussion. Thus the benefit of the fast convergence of Newton's method is restored. In fact based on insights derived from this Newton step interpretation, it appears that the performance penalty for the CE approximation is often small. At the same time the $\ell$-step lookahead minimization involves only one stochastic step, the first one, and hence potentially a much ``thinner" lookahead graph, than an $\ell$-step minimization that does not involve any CE-type  approximations.


\section{MPC and Adaptive Control}

Our discussion so far dealt with problems with a known mathematical model, i.e., one where the system equation, cost function, control constraints, and probability distributions of disturbances are perfectly known. The mathematical model may be available through explicit formulas and assumptions, or through a computer program that can emulate all of the mathematical operations involved in the model, including Monte Carlo simulation for the calculation of expected values.\footnote{The term ``model-free"  is often used to describe the latter situation, but in reality there is a mathematical model that is hidden inside the simulator, so the ideas of present section apply in principle.} 
In practice, however, it is common that the system
parameters are either not known exactly or can change over time, and this introduces potentially enormous complications.\footnote{The difficulties introduced by a changing environment complicate the balance between off-line training and on-line play. It is worth keeping in mind that as much as learning to play high quality chess is a great challenge, the rules of play are stable; they do not change unpredictably in the middle of a game! Problems with changing system parameters can be far more challenging!}

Let us also note that unknown problem environments are an integral part of the artificial intelligence view of RL. In particular, to quote from the popular book by Sutton and Barto  [SuB18],  ``learning from interaction with the environment is  a foundational idea underlying nearly all theories of learning and intelligence"  while RL is described as ``a computational approach to learning from interaction with the environment." 
The idea of learning from interaction with the environment is often connected with the idea of exploring the environment to identify its characteristics. 

In control theory this is often viewed as part of the {\it system identification} methodology, which aims to construct mathematical models of dynamic systems. The system identification process is often combined with the control process to deal with unknown or changing problem parameters, in a framework that is sometimes called {\it dual control\/}. \footnote{The dual control  framework was introduced in a series of papers by Feldbaum, starting in 1960 with  [Fel60]. These papers emphasized the division of effort between system estimation and control, now more commonly referred to as the {\it exploration-exploitation tradeoff\/}. In the last paper of the series [Fel63], Feldbaum prophetically concluded as follows: ``At the present time, the most important problem for the immediate future is the development of approximate solution methods for dual control theory problems, the formulation of sub-optimal strategies, the determination of the numerical value of risk in quasi-optimal systems and its comparison with the value of risk in existing systems."} This is one of the most challenging areas of stochastic optimal and suboptimal control, and has been studied intensively since the early 1960s, with several textbooks and research monographs written: Astr\"om and Wittenmark [AsW94],  Astr\"om and Hagglund [AsH06], Bodson [Bod20], Goodwin  and Sin [GoS84], Ioannou and Sun [IoS96], Jiang  and Jiang [JiJ17], Krstic, Kanellakopoulos, and Kokotovic [KKK95], Kumar and Vara\-iya [KuV86],  Liu, et al.\ [LWW17], Lavretsky and Wise [LaW13], Narendra and Annaswamy [NaA12], Sastry and Bodson [SaB11], Slotine and Li [SlL91], and Vrabie, Vamvouda\-kis, and Lewis [VVL13]. These books describe a vast array of methods spanning 60 years, and ranging from adaptive and model-free approaches, to self-tuning regulators, to simultaneous or sequential control and identification, to time series models, to extremum-seeking methods, to simulation-based RL techniques, etc. 

In this section, we will briefly review some of the most commonly used approaches for dealing with unknown parameters, such as robust control, PID control, and  indirect adaptive control. We will also suggest a simplified version of indirect adaptive control that uses rollout (possibly truncated and supplemented with terminal cost approximation)  in place of policy reoptimization.


\subsection{Robust Control}

Given a controller design that has been obtained assuming a nominal DP problem model, one possibility is to simply ignore changes in problem parameters. We may then try to investigate the performance of the current design for a suitable range of problem parameter values, and ensure that it is adequate for the entire range. This is sometimes called a {\it robust controller design\/}. 

A simple time-honored robust/adaptive control approach for continuous-state problems is {\it PID (Proporti\-onal-Integral-Derivative) control\/}.\footnote{According to Wikipedia, ``a formal control law for what we now call PID or three-term control was first developed using theoretical analysis, by Russian American engineer Nicolas Minorsky" in 1922 [Min22].} The control theory and practice literature contains extensive accounts.
In particular, PID control aims to maintain the output of a single-input single-output dynamic system around a set point or to follow a given trajectory, as the system parameters change within a relatively broad range. In its simplest form, the PID controller is parametrized by three scalar parameters, which may be determined by a variety of methods, some of them manual/heuristic. PID control is used widely and with success, although its range of application is mainly restricted to single-input, single-output continuous-state control systems.

\subsection{Dealing with Unknown Parameters Through System Identification and Reoptimization - On-Line Replanning}

\begin{figure}
\begin{center}
\includegraphics[width=6.0cm]{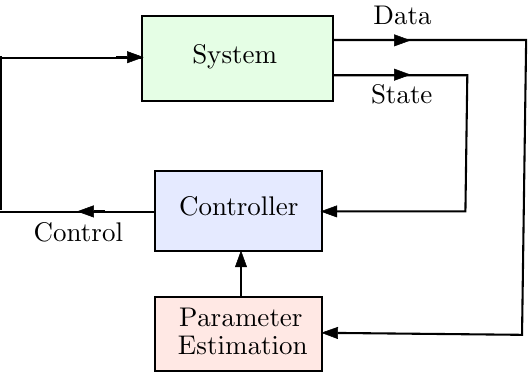}    
\caption{Schematic illustration of on-line replanning: the concurrent parameter estimation and system control. The system parameters are estimated on-line and the estimates are periodically passed on to the controller.}\label{figdualcontrol}
\end{center}
\end{figure}

In PID control, no attempt is made to maintain a mathematical model and to track unknown model parameters as they change. 
A more ambitious form of suboptimal control  is to separate the control process into two phases, a {\it system 
identification phase} and a {\it control phase\/}.  In the first phase the
unknown parameters are estimated, while the control takes no account of the
interim results of estimation.  The final parameter estimates from the
first phase are then used to implement an optimal or suboptimal policy in the second
phase.  This alternation of estimation and control phases may be repeated
several times during any system run in order to take into account subsequent
changes of the parameters.
Moreover, it is not necessary to introduce a hard separation between the identification and the control phases. They may be going on simultaneously, with new parameter estimates being introduced into the control process, whenever this is thought to be desirable; see Fig.\ \ref{figdualcontrol}. This approach is often called  {\it on-line replanning} and is generally known as {\it indirect adaptive control} in the adaptive control literature, see e.g., Astr\"om and Wittenmark [AsW94].

Unfortunately, there is still another difficulty with this type of on-line replanning: it may be hard to recompute an optimal or near-optimal policy on-line, using a newly identified system model. In particular, it may be impossible to use time-consuming methods that involve for example the training of a neural network or discrete/integer control constraints. A simpler possibility is to use approximation in value space that uses rollout with some kind of robust base policy. We discuss this approach next.\footnote{Still another possibility is to deal with this difficulty by precomputation. In particular, assume that the set of problem parameters may take a known finite set of values (for example each set of parameter values may correspond to a distinct maneuver of a vehicle, motion of a robotic arm, flying regime of an aircraft, etc). Then we may precompute a separate controller for each of these values. Once the control scheme detects a change in problem parameters, it switches to the corresponding predesigned current controller. This is sometimes called a {\it multiple model control design} or {\it gain scheduling\/}, and has been applied with success in various settings over the years.}

\subsection{Adaptive Control by Rollout}

We will now consider dealing with unknown or changing parameters by means of an approximate form of on-line replanning that is based on rollout. 
Let us assume that some problem parameters change and the current controller becomes aware of the change ``instantly" (i.e., very quickly, before the next control needs to be applied). The method by which the problem parameters are recalculated or become known is immaterial for the purposes of the following discussion. It may involve a limited form of parameter estimation, whereby the unknown parameters are ``tracked" by data collection over a few time stages; or it may involve new features of the control environment, such as a changing number of servers and/or tasks in a service system. 

\begin{figure}
\begin{center}
\includegraphics[width=8.0cm]{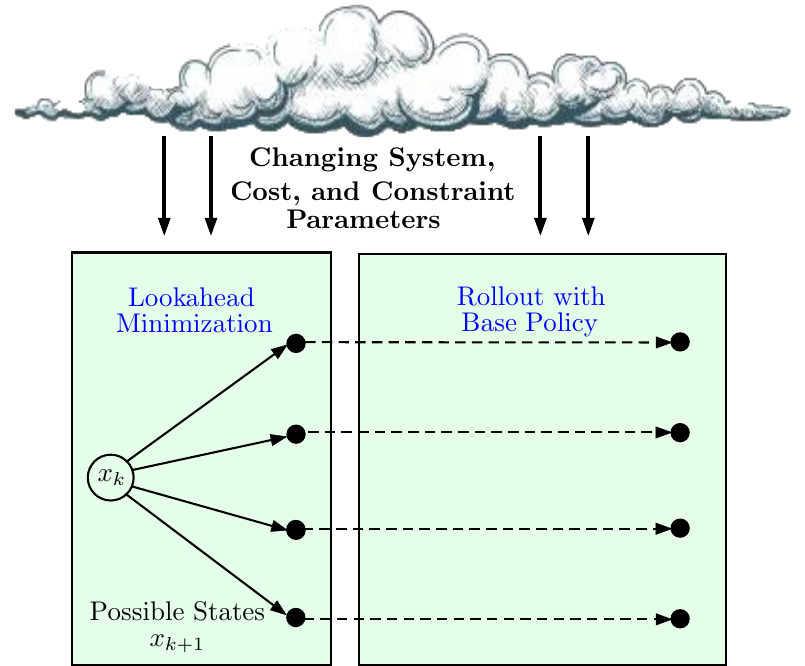}    
\caption{Schematic illustration of adaptive control by rollout.  One-step lookahead is followed by simulation with the base policy, which stays fixed. The system, cost, and constraint parameters are changing over time, and the most recent values are incorporated into the lookahead minimization and rollout operations. For the discussion in this section, we may assume that all the changing parameter information is provided by some computation and sensor ``cloud" that is beyond our control. The base policy may also be revised based on various criteria. Moreover the lookahead minimization may involve multiple steps, while the rollout may be truncated.}
\label{figAdaptRollout}
\end{center}
\end{figure}

We thus assume away/ignore issues of parameter estimation, and focus on revising the controller by on-line replanning based on the newly obtained parameters. This revision may be based on any suboptimal method, but  rollout with the current policy used as the base policy is particularly attractive. Here the advantage of rollout is that it is simple and reliable. In particular, it does not require a complicated training procedure to revise the current policy, based for example on the use of neural networks or other approximation architectures, so {\it no new policy is explicitly computed in response to the parameter changes\/}. Instead the current policy is used as the base policy for (possibly truncated) rollout, and the available controls at the current state are compared by a one-step or mutistep minimization, with cost function approximation provided by the base policy (cf.\ Fig.\ \ref{figAdaptRollout}).

Note  that {\it over time the base policy may also be revised} (on the basis of an unspecified rationale). In this case the rollout policy will be adjusted both in response to the changed current policy and in response to the changing parameters. This is necessary in particular when the constraints of the problem change.

The principal requirement for using rollout in an adaptive control context is that the rollout control computation should be fast enough to be performed between successive control applications. Note, however,  that accelerated/truncated versions of rollout, as well as parallel computation, can be used to meet this time constraint. 

We will now present a one-dimensional linear-quadratic example of on-line replanning involving the use of rollout. The purpose of the example is 
to illustrate how rollout with a policy that is optimal for a nominal set of problem parameters works well when the parameters change from their nominal values. This property is not practically useful in linear-quadratic problems because when the parameters change, it is possible to calculate the new optimal policy in closed form, but it is indicative of the performance robustness of rollout in other contexts. 

Consider the deterministic one-dimensional undiscounted infinite horizon linear-quadratic problem involving the linear system
$$x_{k+1}=x_k+bu_k,$$
and the quadratic cost function
$$\lim_{N\tends\infty}\sum_{k=0}^{N-1}(x_k^2+ru_k^2).$$
The optimal cost function is given by
$$J^*(x)=K^*x^2,$$
where $K^*$ is solves the Riccati equation
$$K={rK \over r+b^2 K }+1.$$
The optimal policy has the form
$$\m^*(x)=L^*x,$$
where
$$L^*=-{ bK^*\over r+b^2K^*}.$$

We will consider the nominal problem parameters $b=2$ and $r=0.5$. We can then verify that for these parameters, the corresponding optimal cost and optimal policy coefficients are
$$K={2+\sqrt{6}\over 4},\qquad L=-{2+\sqrt{6}\over 5+2\sqrt{6}}.$$
We will now consider changes of the values of $b$ and $r$ while keeping $L$ constant, and we will compare the quadratic cost coefficient of the following  cost functions as $b$ and $r$ vary:

\begin{itemize}
\item[(a)] The optimal cost function $K^*x^2$.
\item[(b)] The cost function $K_Lx^2$ that corresponds to the base policy 
$\m_L(x)=Lx.$
From our earlier discussion, we have
$$K_L={1+rL^2\over 1-(1+bL)^2}.$$
\item[(c)] The cost function $\tl K_Lx^2$ that corresponds to the rollout policy 
$\tl\m_L(x)=\tl L x,$ 
obtained by using the policy $\m_L$ as base policy. Using the formulas given earlier, we have
$$\tl L=-{bK_L\over r+b^2K_L},$$
and
$$\tl K_L={1+r\tl L^2\over 1-(1+b\tl L)^2}.$$
\end{itemize}

 \begin{figure}
\begin{center}
\includegraphics[width=8.2cm]{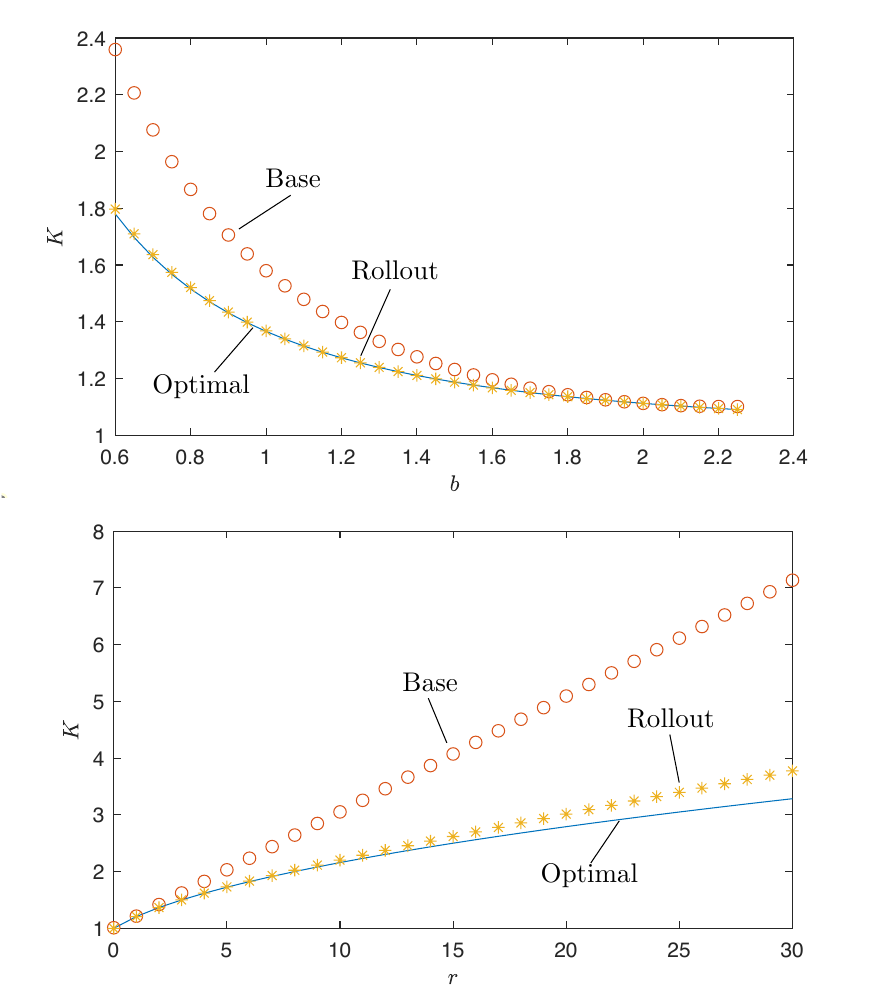}    
\caption{Illustration of adaptive control by rollout under changing problem parameters. The quadratic cost coefficients $K^*$ (optimal, denoted by solid line), $K_L$ (base policy, denoted by circles), and $\tl K_L$ (rollout policy, denoted by asterisks) for the two separate cases where $r=0.5$ and $b$ varies, and $b=2$ and $r$ varies. The value of $L$ is fixed at the value that is optimal for $b=2$ and $r=0.5$
}
\label{figadaptive}
\end{center}
\end{figure}

Figure \ref{figadaptive} shows the coefficients $K^*$, $K_L$, and $\tl K_L$ for a range of values of  $r$ and $b$. As predicted by the cost improvement property of rollout, we have 
$$K^*\le\tl K_L\le K_L.$$
The difference $K_L-K^*$ is indicative of the robustness of the policy $\m_L$, i.e., the performance loss incurred by ignoring the values of $b$ and $r$, and continuing to use the policy $\m_L$, which is optimal for the nominal values $b=2$ and $r=0.5$, but suboptimal for other values of $b$ and $r$. The difference $\tl K_L-K^*$ is indicative of the performance loss due to using on-line replanning by rollout rather than using optimal replanning. Finally, the difference $K_L-\tl K_L$ is indicative of the performance improvement due to on-line replanning using rollout rather than keeping the policy $\m_L$ unchanged. 

It can be seen that the rollout policy performance is very close to the one of the exactly reoptimized policy, while the base policy yields much worse performance. 
This is a consequence of the superlinear convergence rate of Newton's method that underlies rollout:
$$\lim_{J\to J^*}{\skew5\tl J(x)-J^*(x)\over J(x)-J^*(x)}=0,$$
where  for a given initial state $x$, $\skew5\tl J(x)$ is the rollout cost, $J^*(x)$ is the optimal cost, and $J(x)$ is the base policy cost.


\section{Concluding Remarks}

We have argued that the connections between the  MPC and RL fields  are strong, and that the most successful design architectures of the two fields share important characteristics, which relate  to Newton's method. Indeed, in the author's view, a principal theoretical reason for the successes of the two fields  is the off-line training/on-line play synergism that rests upon the mathematical foundations of Newton's method.

Still the cultures of MPC and RL have different starting points and have grown in different directions. One of the primary reasons is the preference for continuous state and control spaces in MPC, which stems from the classical control theory tradition. Thus stability and safety/reachability issues have been of paramount importance in MPC, but they are hardly ever considered in RL. The main reason is that stability does not arise mathematically or practically in the discrete state and control contexts of games, Markovian decision problems, and more recently large language models that are favored in RL. At the same time,  the ideas of learning from data are not part of the control theory tradition, and they have only been addressed relatively recently in a systematic way.

The framework that we have presented in this paper also aims to support a trend of increased use of machine learning methods in MPC. The fact that at their foundation, MPC and RL share important principles suggests  that this trend will continue and accelerate in the future.


\def\refspace{\par\noindent}
\def\ref{\par\noindent}

\section{References}
  \vspace{-5pt}

  \ref[ABQ99] Allgower, F., Badgwell, T.\ A., Qin, J.\ S., Rawlings, J.\ B., and Wright, S.\ J., 1999.\ ``Nonlinear Predictive Control and Moving Horizon Estimation - An Introductory Overview,"Advances in Control: Highlights of ECC'99, pp.\ 391-449.
  
\ref[AGH19] Andersson, J.\ A., Gillis, J., Horn, G., Rawlings, J.\ B., and Diehl, M., 2019.\ ``CasADi: A Software Framework for Nonlinear Optimization and Optimal Control," Math.\ Programming Computation, Vol.\ 11, pp.\ 1-36.
  
\ref [Abr90] Abramson, B., 1990.\ ``Expected-Outcome: A General Model of Static
Evaluation," IEEE Trans.\ on Pattern Analysis and Machine Intelligence,
Vol.\ 12, pp.\ 182-193.

\ref[AsH06] Astr\"om, K.\ J., and Hagglund, T.,  2006.\ Advanced PID Control, Instrument Society of America, Research Triangle Park, NC.

\ref [AsW94]  Astr\"om, K.\ J., and Wittenmark, B., 1994.\ Adaptive Control, 2nd Ed., 
Prentice-Hall, Englewood Cliffs, NJ.

\ref[BBB22] Bhambri, S., Bhattacharjee, A., and Bertsekas, D.\ P., 2022.\ ``Reinforcement Learning Methods for Wordle: A POMDP/Adaptive Control Approach," arXiv:2211.10298.

\ref[BBM17] Borrelli, F., Bemporad, A., and Morari, M., 2017.\ Predictive Control for Linear and Hybrid Systems, Cambridge Univ.\ Press, Cambridge, UK.

\ref[BDL09] Bolte, J., Daniilidis, A., and Lewis, A., 2009.\ ``Tame Functions are Semismooth," Math.\ Programming, Vol.\ 117, pp.\ 5-19.

\ref[BGH22] Brunke, L., Greeff, M., Hall, A.\ W., Yuan, Z., Zhou, S., Panerati, J., and Schoellig, A.\ P., 2022.\ ``Safe Learning in Robotics: From Learning-Based Control to Safe Reinforcement Learning," Annual Review of Control, Robotics, and Autonomous Systems, Vol.\ 5, pp.\ 411-444.

\ref[BKB20] Bhattacharya, S., Kailas, S., Badyal, S., Gil, S., and Bertsekas, D.\ P., 2020.\ ``Multiagent Rollout and Policy Iteration for POMDP with Application to Multi-Robot Repair Problems," in Proc.\ of Conference on Robot Learning (CoRL); also arXiv preprint, arXiv:2011.04222.

\refspace[BTW97] Bertsekas, D.\ P., Tsitsiklis, J.\ N., and Wu, C., 1997.\ ``Rollout Algorithms for
Combinatorial Optimization,'' Heuristics, Vol.\ 3, pp.\ 245-262.

\ref[BeC99] Bertsekas, D.\ P., and  Casta\~non, D.\ A., 1999.\ ``Rollout Algorithms for
Stochastic Scheduling Problems," Heuristics, Vol.\ 5, pp.\ 89-108. 

\ref[BeK65] Bellman, R., and Kalaba, R.\ E., 1965.\ Quasilinearization and Nonlinear Boundary-Value Problems, Elsevier, NY.

\ref[BeM99] Bemporad, A., and Morari, M., 1999.\ ``Control of Systems Integrating Logic, Dynamics, and Constraints," Automatica, Vol.\ 35, pp.\ 407-427.

\ref[BeP21] Bemporad, A., and Piga, D., 2021.\ ``Global Optimization Based on Active Preference Learning with Radial Basis Functions," Machine Learning, Vol.\ 110, pp.\ 417-448.

\ref [BeR71] Bertsekas, D.\ P., and Rhodes, I.\ B., 1971.\ ``On the Minimax
Reachability of Target Sets and Target Tubes," Automatica,
Vol.\ 7, pp.\ 233-247.

\ref [BeS78]  Bertsekas, D.\ P., and Shreve, S.\ E., 1978.\  Stochastic Optimal
Control:  The Discrete Time Case, Academic Press, NY.;
republished by Athena Scientific, Belmont, MA, 1996 (can be  downloaded from
the author's website).

\ref [BeT96]  Bertsekas, D.\ P., and Tsitsiklis, J.\ N., 1996.\ Neuro-Dynamic Programming,
Athena Scientific, Belmont, MA.

\ref [Ber71] Bertsekas, D.\ P., 1971.\ ``Control of Uncertain Systems With a
Set-Member\-ship Description of the Uncertainty," Ph.D.\ Dissertation, 
Massachusetts
Institute of Technology, Cambridge, MA (can be  downloaded from
the author's website).

\ref [Ber72] Bertsekas, D.\ P., 1972.\  ``Infinite Time Reachability of State
Space Regions by Using Feedback Control," IEEE Trans.\ Aut.\ Control, Vol.
AC-17, pp.\ 604-613.

\ref [Ber77] Bertsekas, D.\ P., 1977.\  ``Monotone Mappings with Application in
Dynamic Programming," SIAM J.\ on Control and Opt., Vol.\ 15, pp.\
438-464.

\ref[Ber97] Bertsekas, D.\ P., 1997.\ ``Differential Training of Rollout Policies,"
Proc. of the 35th Allerton Conference on Communication, Control, and Computing,
Allerton, Ill.

\ref[Ber05a] Bertsekas, D.\ P., 2005.\ ``Dynamic Programming and Suboptimal Control: A Survey from ADP to MPC," European J.\ of Control, Vol.\ 11, pp.\ 310-334.

\ref[Ber05b] Bertsekas, D.\ P., 2005.\ ``Rollout Algorithms for Constrained Dynamic Programming," Lab. for Information and Decision Systems Report LIDS-P-2646, MIT.

\refspace[Ber17a] Bertsekas, D.\ P., 2017.\ Dynamic Programming and Optimal Control,  Vol.\ I, Athena Scientific, Belmont, MA.

\refspace[Ber17b] Bertsekas, D.\ P., 2017.\ ``Value and Policy Iteration in Deterministic Optimal Control and Adaptive Dynamic Programming,"  IEEE Transactions on Neural Networks and Learning Systems, Vol.\ 28, pp.\ 500-509.

\refspace[Ber19]   Bertsekas, D.\ P., 2019.\ Reinforcement Learning and Optimal Control, Athena Scientific, Belmont, MA.

\refspace[Ber20]   Bertsekas, D.\ P., 2020.\ Rollout, Policy Iteration, and Distributed Reinforcement Learning,  Athena Scientific, Belmont, MA.

\refspace[Ber21a] Bertsekas, D.\ P., 2021.\ ``Multiagent Reinforcement Learning: Rollout and Policy Iteration," IEEE/CAA Journal of Automatica Sinica, Vol.\ 8, pp. 249-271.

\ref[Ber21b] Bertsekas, D.\ P., 2021.\ ``Distributed Asynchronous Policy Iteration for Sequential Zero-Sum Games and Minimax Control," arXiv:2107.10406

\refspace[Ber22a]  Bertsekas, D.\ P., 2022.\ ``Lessons from AlphaZero for Optimal, Model Predictive, and Adaptive Control," Athena Scientific, Belmont, MA (can be  downloaded from the author's website).

\refspace[Ber22b]  Bertsekas, D.\ P., 2022.\ Abstract Dynamic Programming, 3rd Ed., Athena Scientific, Belmont, MA  (can be  downloaded from the author's website).

\refspace[Ber22c] Bertsekas, D.\ P., 2022.\ ``Newton's Method for Reinforcement Learning and Model Predictive Control," Results in Control and Optimization, Vol.\ 7, pp.\ 100-121.

\refspace[Ber23]   Bertsekas, D.\ P., 2023.\ ``A Course in Reinforcement Learning," Athena Scientific, 2023 (can be  downloaded from the author's website).

\ref [Bla99] Blanchini, F., 1999.\ ``Set Invariance in Control -- A Survey," 
Automatica, Vol.\ 35, pp.\ 1747-1768.

\ref[Bod20] Bodson, M., 2020.\  Adaptive Estimation and Control, Independently Published.

\ref[CFM20] Chen, S., Fazlyab, M., Morari, M., Pappas, G.\ J., and Preciado, V.\ M., 2020.\ ``Learning Lyapunov Functions for Piecewise Affine Systems with Neural Network Controllers," arXiv preprint arXiv:2008.06546.

\ref[CLD19] Coulson, J., Lygeros, J., and Dorfler, F., 2019.\ ``Data-Enabled Predictive Control: In the Shallows of the DeePC," 18th European Control Conference, pp.\ 307-312.

\ref[CLL23] Choi, J.\ J., Lee, D., Li, B., How, J.\ P., Sreenath, K., Herbert, S.\ L., and Tomlin, C.\ J., 2023.\ ``A Forward Reachability Perspective on Robust Control Invariance and Discount Factors in Reachability Analysis," arXiv preprint arXiv:2310.17180.

\ref[CMT87a] Clarke, D.\ W., Mohtadi, C., and Tuffs, P.\ S., 1987.\ ``Generalized Predictive Control - Part I.\ The Basic Algorithm," Automatica, Vol.\ 23, pp.\ 137-148.

\ref[CMT87b] Clarke, D.\ W., Mohtadi, C., and Tuffs, P.\ S., 1987.\ ``Generalized Predictive Control - Part II," Automatica, Vol.\ 23, pp.\ 149-160.

\ref[CWA22] Chen, S.\ W., Wang, T., Atanasov, N., Kumar, V., and Morari, M., 2022.\ ``Large Scale Model Predictive Control with Neural Networks and Primal Active Sets," Automatica, Vol.\ 135.

\ref[CaB07] Camacho, E.\ F., and Bordons, C., 2007.\ Model Predictive Control, 2nd
Ed., Springer, New York, NY.

\ref[DFH09] Diehl, M., Ferreau, H.\ J., and Haverbeke, N., 2009.\ ``Efficient Numerical Methods for Nonlinear MPC and Moving Horizon Estimation," in Nonlinear Model Predictive Control: Towards New Challenging Applications, by L.\ Magni, D.\ M.\ Raimondo, F.\ Allgower (eds.), Springer, pp.\ 391-417.

\ref[DMS98] De Nicolao, G., Magni, L., and Scattolini, R., 1998.\
``Stabilizing Receding-Horizon Control of Nonlinear
Time-Varying Systems," IEEE Transactions on Aut.\
Control, Vol.\ 43, pp.\ 1030-1036.

\ref[DuM23] Duan, Y., and Wainwright, M.J., 2023.\ ``A Finite-Sample Analysis of Multi-Step Temporal Difference Estimates," in Learning for Dynamics and Control Conference, N.\ Matni, M.\ Morari, G.\ J.\ Pappas (eds.), Proc.\ of Machine Learning Research, pp.\ 612-624.

\ref [FHS09] Feitzinger, F., Hylla, T., and Sachs, E.\ W., 2009.\ ``Inexact Kleinman-Newton Method for Riccati Equations," SIAM Journal on Matrix Analysis and Applications, Vol.\ 3, pp.\ 272-288.

\ref[FXB22] Fu, A., Xing, L., and Boyd, S., 2022.\ ``Operator Splitting for Adaptive Radiation Therapy with Nonlinear Health Dynamics," Optimization Methods and Software, Vol.\ 37, pp.\ 2300-2323.

\ref[FIA03] Findeisen, R., Imsland, L., Allgower, F., and Foss, B.A., 2003.\ ``State and Output Feedback Nonlinear Model Predictive Control: An Overview," European Journal of Control, Vol.\ 9, pp.\ 190-206.

\ref[FaP03] Facchinei, F., and Pang, J.-S., 2003.\ Finite-Dimensional Variational Inequalities and Complementarity Problems,
Vols I and II, Springer, NY.

\ref[Fel60] Feldbaum, A.\ A., 1960.\ ``Dual Control Theory," Automation and Remote Control, Vol.\ 21, pp.\ 874-1039.

\ref[Fel63] Feldbaum, A.\ A., 1963.\ ``Dual Control Theory Problems," IFAC Proceedings, pp.\ 541-550.

\ref[GFA11] Gonzalez, R., Fiacchini, M., Alamo, T., Guzman, J.\ L., and Rodriguez, F., 2011.\ ``Online Robust Tube-Based MPC for Time-Varying Systems: A Practical Approach," International Journal of Control, Vol.\ 84, pp.\ 1157-1170.

\ref [GPG22] Garces, D., Bhattacharya, S., Gil, G., and Bertsekas, D., 2022.\ ``Multiagent Reinforcement Learning for Autonomous Routing and Pickup Problem with Adaptation to Variable Demand," arXiv preprint arXiv:2211.14983.

\ref[GSD06] Goodwin, G., Seron, M.\ M., and De Dona, J.\ A., 2006.\ Constrained Control and Estimation: An Optimisation Approach, Springer, NY.

\ref [GoS84] Goodwin, G.\ C., and Sin, K.\ S.\ S., 1984.\  Adaptive Filtering,
Prediction, and Control, Prentice-Hall, Englewood Cliffs, N.\ J.

\ref[GrZ19] Gros, S., and Zanon, M., 2019.\ ``Data-Driven Economic NMPC Using Reinforcement Learning," IEEE Trans.\ on Aut.\ Control, Vol.\ 65, pp.\ 636-648.

\ref[GrZ22] Gros, S., and Zanon, M., 2022.\ ``Learning for MPC with Stability and Safety Guarantees," Automatica, Vol.\ 146, pp.\ 110598.

\ref[HWM20] Hewing, L., Wabersich, K.\ P., Menner, M., and Zeilinger, M.\ N., 2020.\ ``Learning-Based Model Predictive Control: Toward Safe Learning in Control," Annual Review of Control, Robotics, and Autonomous Systems, Vol.\ 3, pp.\ 269-296.

\ref[Hew71] Hewer, G., 1971.\ ``An Iterative Technique for the Computation of the Steady State Gains for the Discrete Optimal Regulator," IEEE Trans.\ on Aut.\ Control, Vol.\ 16, pp.\ 382-384. 

\ref[Hyl11] Hylla, T., 2011.\ Extension of Inexact Kleinman-Newton Methods to a General Monotonicity Preserving Convergence Theory, PhD Thesis, Univ.\ of Trier.

\ref [IoS96] Ioannou, P.\ A., and Sun, J., 1996.\ Robust Adaptive Control, Prentice-Hall, Englewood
Cliffs, N.\ J.

\ref[ItK03] Ito, K., and Kunisch, K., 2003.\ ``Semi-Smooth Newton Methods for Variational Inequalities of the First Kind,"  Mathematical Modelling and Numerical Analysis, Vol.\ 37, pp.\ 41-62.

\ref[JiJ17] Jiang, Y., and Jiang, Z.\ P., 2017.\ Robust Adaptive Dynamic Programming, J.\ Wiley, NY.

\ref[Jos79] Josephy, N.\ H., 1979. ``Newton's Method for Generalized Equations," Wisconsin Univ-Madison, Mathematics Research Center Report No.\ 1965.

\ref [KGB82] Kimemia, J., Gershwin, S.\ B., and Bertsekas, D.\ P., 1982.\ 
``Computation of Production Control Policies by a Dynamic Programming
Technique," in Analysis and Optimization of Systems, A.\ Bensoussan and J.\ L.
Lions (eds.), Springer, N.\ Y., pp.\ 243-269.

\ref [KKK95] 
Krstic, M., Kanellakopoulos, I., Kokotovic, P., 1995.\ Nonlinear and Adaptive Control
Design, J.\ Wiley, NY.

\ref[KRW21] Kumar, P., Rawlings, J.\ B., and Wright, S.\ J., 2021.\ ``Industrial, Large-Scale Model Predictive Control with Structured Neural Networks," Computers and Chemical Engineering, Vol.\ 150.

\ref [KeG88] Keerthi, S.\ S., and Gilbert, E.\ G., 1988.\ ``Optimal, Infinite Horizon
Feedback Laws for a General Class of Constrained Discrete Time Systems: Stability and
Moving-Horizon Approximations," J.\ Optimization Theory Appl., Vo.\ 57, pp.\ 265-293.

\ref [Ker00] Kerrigan, E.\ C., 2000.\ Robust Constraint Satisfaction: Invariant Sets and Predictive Control, PhD.\ Thesis, University of London.

\ref [Kle68] Kleinman, D.\ L., 1968.\  ``On an Iterative Technique for Riccati
Equation Computations," IEEE Trans.\ Aut.\  Control, Vol.\ AC-13, pp.\ 114-115.

\ref[KoC16] Kouvaritakis, B., and Cannon, M., 2016.\ Model Predictive Control: Classical, Robust and Stochastic, Springer, NY.

\ref[KoG98] Kolmanovsky, I., and Gilbert, E.\ G., 1998.\ ``Theory and Computation of Disturbance Invariant Sets for Discrete-Time Linear Systems," Mathematical Problems in Engineering, Vol.\ 4, pp.\ 317-367.

\ref[KoS86] Kojima, M., and Shindo, S., 1986.\ ``Extension of Newton and Quasi-Newton Methods to Systems of $PC^ 1$ Equations," J.\ of the Operations Res. Society of Japan, Vol.\ 29, pp.\ 352-375.

\ref[Kre19] Krener, A.\ J., 2019.\ ``Adaptive Horizon Model Predictive Control and Al'brekht's Method," arXiv preprint arXiv:1904.00053.

\ref [KuV86] Kumar, P.\ R., and Varaiya, P.\ P., 1986.\  Stochastic Systems: 
Estimation, Identification, and Adaptive Control, Prentice-Hall, Englewood
Cliffs, N.\ J.

\ref[Kum88] Kummer, B., 1988.\ ``Newton's Method for Non-Differentiable Functions," Mathematical Research, Vol.\ 45, pp.\ 114-125.

\ref[Kum00] Kummer, B., 2000.\ ``Generalized Newton and NCP-methods: Convergence, Regularity, Actions," Discussiones Mathematicae, Differential Inclusions, Control and Optimization, Vol.\ 2, pp.\ 209-244.

\ref[LFE13] Lucia, S., Finkler, T., and Engell, S., 2013.\ ``Multi-Stage
Nonlinear Model Predictive Control Applied to a Semi-Batch
Polymerization Reactor Under Uncertainty," Journal of
Process Control, Vol.\ 23, pp.\ 1306-1319.

\ref[LHK18] Liao-McPherson, D., Huang, M., and Kolmanovsky, I., 2018.\ ``A Regularized and Smoothed Fischer?Burmeister Method for Quadratic Programming with Applications to Model Predictive Control," IEEE Trans.\ on Automatic Control, Vol.\ 64, pp.\ 2937-2944.

\ref[LJM21] Li, Y., Johansson, K.\ H., Martensson, J., and Bertsekas, D.\ P., 2021. ``Data-Driven Rollout for Deterministic Optimal Control," arXiv preprint arXiv:\-2105.03116.

\ref[LKL23]  Li, Y., Karapetyan, A., Lygeros, J., Johansson, K.
 H., and Martensson, J., 2023.\ ``Performance Bounds of Model Predictive Control for Unconstrained and Constrained Linear Quadratic Problems and Beyond," IFAC-Papers On Line, Vol.\ 56, pp.\  8464-8469.

\ref[LWW17] Liu, D., Wei, Q., Wang, D., Yang, X., and Li, H., 2017.\ Adaptive Dynamic Programming with Applications in Optimal Control, Springer, Berlin.

\ref[LaW13] Lavretsky, E., and Wise, K., 2013.\ Robust and Adaptive Control with Aerospace Applications, Springer.

\ref[Li23] Li, Y., 2023.\ Approximate Methods of Optimal Control via Dynamic Programming Models, PhD Thesis, Royal Institute of Technology, Stockholm.

\ref[LiB24] Li, Y., and Bertsekas, D., 2024.\ ``Most Likely Sequence Generation for $n$-Grams, Transformers, HMMs, and Markov Chains, by Using Rollout Algorithms," arXiv:2403.15465.

\ref[MBS23] Moreno-Mora, F., Beckenbach, L., and Streif, S., 2023.\ ``Predictive Control with Learning-Based Terminal Costs Using Approximate Value Iteration," IFAC-Papers On Line, Vol.\ 56, pp.\ 3874-3879.

\ref[MDM01] Magni, L., De Nicolao, G., Magnani, L., and Scattolini, R., 2001.\ ``A Stabilizing Model-Based Predictive Control Algorithm for Nonlinear Systems," Automatica, Vol.\ 37, pp.\ 1351-1362.

\ref[MGQ20] Mittal, M., Gallieri, M., Quaglino, A., Salehian, S., and Koutnik, J., 2020.\ ``Neural Lyapunov Model Predictive Control: Learning Safe Global Controllers from Suboptimal Examples," arXiv preprint arXiv:2002.10451.

\ref[MDT22] Mukherjee, S., Drgona, J., Tuor, A., Halappanavar, M., and Vrabie, D., 2022.\ Neural Lyapunov Differentiable Predictive Control," 2022 IEEE 61st Conference on Decision and Control, pp.\ 2097-2104. 

\ref[MJR22] Mania, H., Jordan, M.\ I., and Recht, B., 2022.\ ``Active Learning for Nonlinear System Identification with Guarantees," J.\ of Machine Learning Research, Vol.\ 23, pp.\ 1-30.

\ref[MLW24] Musunuru, P., Li, Y., Weber, J., and Bertsekas, D., ``An Approximate Dynamic Programming Framework for Occlusion-Robust Multi-Object Tracking," ArXiv Preprint arXiv:2405.15137, May 2024.

\ref[MRR00] Mayne, D., Rawlings, J.\ B., Rao, C.\ V., and Scokaert, P.\ O.\ M., 2000.\ ``Constrained Model Predictive Control: Stability and Optimality," Automatica, Vol.\ 36, pp.\ 789-814.

\ref[MaM88] Mayne, D.\ Q., and Michalska, H., 1988.\ ``Receding Horizon Control of Nonlinear Systems," Proc.\ of the 27th IEEE Conf.\ on Decision and Control, pp.\ 464-465.

\ref[MaS04] Magni, L., and Scattolini, R., 2004.\ ``Stabilizing Model
Predictive Control of Nonlinear Continuous Time Systems,"
Annual Reviews in Control, Vol.\ 28, pp.\ 1-11.

\ref [May14] Mayne, D.\ Q., 2014.\ ``Model Predictive Control: Recent Developments and Future Promise," Automatica, Vol.\ 50, pp.\ 2967-2986.

{\ref[Min22]  Minorsky, N., 1922.\ ``Directional Stability of Automatically Steered Bodies," J.\ Amer.\ Soc.\ Naval Eng.,Vol.\ 34, pp.\ 280-309.}

\ref [MoL99] Morari, M., and Lee, J.\ H., 1999.\ ``Model Predictive Control: Past,
Present, and Future," Computers and Chemical Engineering, Vol.\ 23, pp.\ 667-682.

\ref[NaA12] Narendra, K.\ S., and Annaswamy, A.\ M., 2012.\ Stable Adaptive Systems, Courier Corp.

\ref[OSB13] O'Donoghue, B., Stathopoulos, G., and Boyd, S., 2013.\ ``A Splitting Method for Optimal Control," IEEE Trans.\ on Control Systems Technology, Vol.\ 21, pp.\ 2432-2442.

\ref[Pan90] Pang, J.\ S., 1990.\ ``Newton's Method for B-Differentiable Equations,"  Math.\ of Operations Res., Vol.\ 15, pp.\ 311-341.

\ref[PoA69] Pollatschek, M.\ A. and Avi-Itzhak, B., 1969.\ ``Algorithms for Stochastic Games with Geometrical Interpretation," Management Science,  Vol.\ 15, pp. 399-415.

\ref [PuB78] Puterman, M.\ L., and Brumelle, S.\ L., 1978.\  ``The Analytic Theory of
Policy Iteration," in Dynamic Programming and Its Applications, M.\ L.\ Puterman
(ed.), Academic Press, NY.

\ref [PuB79] Puterman, M.\ L., and Brumelle, S.\ L., 1979.\  ``On the Convergence of Policy Iteration in Stationary Dynamic Programming," Math.\ of Operations Res., Vol.\ 4, pp.\ 60-69.

\ref[Qi93] Qi, L., 1993.\ ``Convergence Analysis of Some Algorithms for Solving Nonsmooth Equations," Math.\ of Operations Res., Vol.\ 18, pp.\ 227-244.

\ref[QiS93] Qi, L., and Sun, J., 1993.\ ``A Nonsmooth Version of Newton's Method," Math.\ Programming, Vol.\ 58, pp.\ 353-367.

 \ref[RDM24] Ruoss, A., Del\'etang, G., Medapati, S., Grau-Moya, J., Wenliang, L.\ K., Catt, E., Reid, J., and Genewein, T., 2024.\ ``Grandmaster-Level Chess Without Search," arXiv:2402.04494.

\ref[RKM06] Rakovic, S.\ V., Kerrigan, E.\ C., Mayne, D.\ Q., and Lygeros, J., 2006.\ ``Reachability Analysis of Discrete-Time Systems with Disturbances," IEEE Trans.\ on Aut.\ Control, Vol.\ 51, pp.\ 546-561.

 \ref[RMD17] Rawlings, J.\ B., Mayne, D.\ Q., and Diehl, M.\ M., 2017.\ Model Predictive Control: Theory, Computation, and Design, 2nd Ed., Nob Hill Publishing. 
 
 \ref[RaL18] Rakovic, S.\ V., and Levine, W.\ S., eds., 2018.\ Handbook of Model Predictive Control, Springer.
 
 \ref[RaR17] Rawlings, J.\ B., and Risbeck, M.\ J., 2017.\ ``Model Predictive Control with Discrete Actuators: Theory and Application," Automatica, Vol.\ 78, pp.\ 258-265.
 
 \ref[Rec19] Recht, B., 2019.\ ``A Tour of Reinforcement Learning: The View from Continuous Control," Annual Review of Control, Robotics, and Autonomous Systems, Vol.\ 2, pp.\ 253-279.

 \ref[RoB17] Rosolia, U., and Borrelli, F., 2017.\ ``Learning Model Predictive Control for Iterative Tasks. A Data-Driven Control Framework," IEEE Trans.\ on Aut.\ Control, Vol.\ 63, pp.\ 1883-1896.
 
 \ref[Rob80] Robinson, S.\ M., 1980.\ ``Strongly Regular Generalized Equations," Math.\ of Operations Res., Vol.\ 5, pp.\ 43-62.

\ref[Rob88] Robinson, S.\ M., 1988.\ ``Newton's Method for a Class of Nonsmooth Functions,"  Industrial Engineering Working Paper, University of Wisconsin; also in Set-Valued Analysis Vol.\ 2, 1994, pp.\ 291-305.

\ref[Rob11] Robinson, S.\ M., 2011.\ ``A Point-of-Attraction Result for Newton's Method with Point-Based Approximations," Optimization, Vol.\ 60, pp.\ 89-99.

\refspace[SHM16] Silver, D., Huang, A., Maddison, C.\ J., Guez, A., Sifre, L., Van Den Driessche, G., Schrittwieser, J., Antonoglou, I., Panneershelvam, V., Lanctot, M., and Dieleman, S., 2016.\ ``Mastering the Game of Go with Deep Neural Networks and Tree Search," Nature, Vol.\ 529, pp.\ 484-489.

\refspace[SHS17] Silver, D., Hubert, T., Schrittwieser, J., Antono\-glou, I., Lai, M., Guez, A., Lanctot, M., Sifre, L., Kumaran, D., Graepel, T., and Lillicrap, T., 2017.\ ``Mastering Chess and Shogi by Self-Play with a General Reinforcement Learning Algorithm,"  arXiv:1712.01815.

\ref[SKG22] Seel, K., Kordabad, A.\ B., Gros, S., and Gravdahl, J.\ T., 2022.\ ``Convex Neural Network-Based Cost Modifications for Learning Model Predictive Control," IEEE Open Journal of Control Systems, Vol.\ 1, pp. 366-379.

\refspace[SSS17] Silver, D., Schrittwieser, J., Simonyan, K., Antono\-glou, I., Huang, A., Guez, A., Hubert, T., Baker, L., Lai, M., Bolton, A. and Chen, Y., 2017.\ ``Mastering the Game of Go Without Human Knowledge," Nature, Vol.\ 550, pp.\ 354-359.

\ref[SaB11] Sastry, S., and Bodson, M., 2011.\ Adaptive Control: Stability, Convergence and Robustness, Courier Corp.

\ref[SiB22] Silver, D., and Barreto, A., 2022.\ ``Simulation-Based Search," in Proc.\ Int.\ Cong.\ Math, Vol.\ 6, pp.\ 4800-4819.

\ref [SlL91] Slotine, J.-J.\ E., and Li, W., Applied Nonlinear Control,
Prentice-Hall, Englewood Cliffs, N.\ J.

\refspace[TeG96] Tesauro, G., and Galperin, G.\ R., 1996.\ ``On-Line Policy Improvement
Using Monte Carlo Search,'' NIPS, Denver, CO.

\refspace[Tes94] Tesauro, G.\ J., 1994.\ ``TD-Gammon, a Self-Teaching 
Backgammon Program, Achieves Master-Level Play,'' Neural Computation, Vol.\ 6, pp.\ 
215-219.

\refspace[Tes95] Tesauro, G.\ J., 1995.\ ``Temporal Difference Learning and TD-Gammon,'' Communications of the ACM, Vol.\ 38, pp.\ 58-68.

\ref[VVL13] Vrabie, D., Vamvoudakis, K.\ G., and Lewis, F.\ L., 2013.\ Optimal Adaptive Control and Differential Games by Reinforcement Learning Principles,
The Institution of Engineering and Technology, London.

\ref[XDS23] Xie, H., Dai, L., Sun, Z., and Xia, Y., 2023.\ ``Maximal Admissible Disturbance Constraint Set for Tube-Based Model Predictive Control," IEEE Trans.\ on Automatic Control, Vol.\ 68, pp.\ 6773-6780.

\ref [WGP23] Weber, J., Giriyan, D., Parkar, D., Richa, A., and Bertsekas, D., 2023.\ ``Distributed Online Rollout for Multivehicle Routing in Unmapped Environments," arXiv preprint arXiv:2305.11596v1.

\ref[WaB10] Wang, Y., and Boyd, S., 2010.\ ``Fast Model Predictive Control Using Online Optimization," IEEE Trans.\ on Control Systems Tech., Vol.\ 18, pp.\ 267-278.

\ref[Wri19] Wright, S.\ J., 2019.\ ``Efficient Convex Optimization for Linear MPC," Handbook of Model Predictive Control, pp.\ 287-303.

\refspace[YDR04] Yan, X., Diaconis, P., Rusmevichientong, P., and Van Roy, B., 2004.\
``Solitaire: Man Versus Machine,'' Advances in Neural Information
Processing Systems, Vol.\ 17, pp.\ 1553-1560.

\end{document}